\documentclass[apj,twocolumn,twocolappendix,numberedappendix]{openjournal}
\usepackage{newtxtext,newtxmath,enumitem,multirow,ctable,verbatim,graphicx}
\usepackage[T1]{fontenc}
\usepackage[breaklinks,colorlinks,citecolor=blue,urlcolor=blue]{hyperref}

\newcommand{\Alf}{{Alfv\'en}}

\newcommand{\orcidauthor}[3]{\author{\href{http://orcid.org/#1}{#2$^{#3}$}}}

\shorttitle{Cloud Accretion to Magnetized Disks}
\shortauthors{Authors}

\begin{document}
\title{\vspace{-0.8cm}Initial Conditions of Magnetized AGN Accretion Disks from Cloud Capture and Disruption\vspace{-1.5cm}}
\orcidauthor{0000-0003-3729-1684}{Philip F. Hopkins}{1,*}
\affiliation{$^{1}$TAPIR, Mailcode 350-17, California Institute of Technology, Pasadena, CA 91125, USA}
\thanks{$^*$E-mail: \href{mailto:phopkins@caltech.edu}{phopkins@caltech.edu}},

\begin{abstract}
Accretion onto supermassive black holes (BHs) from the interstellar medium (ISM) can occur via capture and circularization of cold gas ``clouds'' (giant molecular clouds, clumps, or overdense structures). Unlike the material in stellar tidal disruptions or X-ray binaries, or classical Bondi accretion, ISM clouds are strongly magnetized (plasma $\beta \equiv P_{\rm thermal}/P_{\rm magnetic} \ll 1$), with supersonic turbulence, and cool rapidly. We derive simple analytic scalings for the ``initial conditions'' of accretion disks formed by capture/disruption of such clouds, in three regimes: (1) low-mass BHs ($M_{\rm BH} \ll M_{\rm cl}$), where turbulent Bondi-Hoyle capture or hierarchical ``leapfrog'' growth via sub-cloud encounters dominates; (2) comparable-mass encounters ($M_{\rm BH} \sim M_{\rm cl}$); and (3) massive BHs tidally disrupting smaller clouds ($M_{\rm BH} \gg M_{\rm cl}$). In all cases, the resulting disk is predicted to be initially (at the circularization radius) toroidally-magnetically-dominated ($|B_{\phi}| \gg |B_{z}|$, $\beta \ll 1$), geometrically thick ($H/R \sim 0.1-1$), and gravitationally stable ($Q \gtrsim 1$), with super-sonic and trans-\Alf{ic} turbulence. These ``initial conditions'' differ qualitatively from those assumed in most classical accretion disk theory, and from  ``chaotic cold accretion'' models, stellar tidal disruption ($\beta \gg 1$), Bondi-Hoyle accretion, or effective-$\alpha$ disk theory (instead resembling a series of Shu-like isothermal collapses). At least initially, the toroidal field should be amplified and sustained by flux-freezing on timescales shorter than Parker instability losses or turbulent resistive decay, in the regimes relevant for AGN. We examine how a cloud mass spectrum drives accretion rate distributions, duty cycles, and time-averaged fueling rates, with the integrated BH mass growth typically dominated by clouds of mass comparable to the BH. Given known cosmological evolution of galactic gas clouds, this naturally suggests a scenario linking (a) inefficient BH growth and wandering BHs in dwarf galaxies, (b) hyper-Eddington growth in little red dots, (c) peak quasar activity in massive gas-rich systems at cosmic noon, and (d) lower-luminosity AGN in gas-poor galaxies at late times. 
\end{abstract}
\maketitle

\section{Introduction}
\label{sec:intro}

Supermassive black hole (BH) accretion powers quasars and active galactic nuclei \citep[AGN;][]{schmidt:1963.qso.redshift,salpeter64}, explaining most of the present BH mass in galaxies \citep{soltan82} with luminosities $L$ and accretion rates $\dot{M}_{\rm BH}$ exceeding $L \gtrsim 10^{46}\,{\rm erg\,s^{-1}}$ \&\ $\dot{M}_{\rm BH} \gtrsim 1\,{\rm M_{\odot}\,yr^{-1}}$  in typical quasars at redshifts $z\gtrsim 2$ to orders of magnitude smaller in low-luminosity AGN \citep[e.g.][and references therein]{shen:bolometric.qlf.update}. 
The gas fueling this accretion comes from the interstellar medium (ISM), beyond the BH radius of influence (BHROI; scales $R_{\rm ROI} \sim G\,M_{\rm BH}/\sigma_{\ast}^{2}\sim $\,a few pc, in terms of the central galaxy potential $\sigma_{\ast}^{2}$ and BH mass $M_{\rm BH}$). This, in turn, means that basic qualitative properties and structure of the accretion are sensitive to the ``outer boundary conditions'' -- i.e.\ how this gas is captured and circularized around the BH \citep{pringle:accretion.review,frank:2002.accretion.book,abramowicz:accretion.theory.review,jafari:2019.mhd.ppd.review}.

There are many regimes where this capture and circularization can be heuristically described by capture and/or disruption of ``clouds'' or cloud-like structures, by which we heuristically mean any over-dense, close-to-self-gravitating gaseous structure in the ISM. The most obvious examples, for AGN, are giant molecular clouds (GMCs), the spectrum of gaseous clouds/clumps/cores, and more general overdense/turbulent structures in the turbulent ISM. It has been recognized that this must form a key part of the accretion process for many AGN, ranging from low-luminosity AGN \citep{ho:2009.llagn.review,hopkins:inflow.analytics,anglesdraft:2017.bh.acc.torques} to the brightest quasars \citep{hopkins:lifetimes.letter,hopkins:2006.qso.all,hopkins:zoom.sims,hopkins:superzoom.overview}. And in an ISM with efficient cooling (e.g.\  galaxies where the nuclear gas is primarily atomic/molecular), even large-scale ``smooth'' flows (e.g.\ bar-directed inflows) must break into individual cold clouds/clumps; likewise even smooth Bondi-like accretion from very hot, virialized X-ray emitting gas can break into cold clumps which dominate the accretion if it becomes thermally unstable \citep{gaspari:2013.chaotic.cold.accretion,gaspari:2015.cca.rotating,gaspari:2017.cca.raining,gaspari:2020.linking,li.bryan:2014.cooling.flow.cold.gas,tremblay:2016.m87.cold.gas}. 

At first, it might seem like the initial dynamics and circularization of such cloud-accretion events, and properties of the accretion disks which form, would be described by decades-old, well-studied analytic models for either the tidal disruption of stars (if the cloud is small compared to the BH) and/or Bondi-Hoyle accretion (if the cloud is large compared to the BH). 
But those historical models make some critical assumptions which are \textit{not} valid for ISM clouds. (1) They assume that the material being accreted is thermal-pressure-dominated, i.e.\ has plasma $\beta_{\rm cl} \equiv P_{\rm thermal} / P_{\rm magnetic} \sim c_{s}^{2} / v_{A}^{2} \gg 1$, with trans or sub-sonic turbulence. And (2) they assume that this material has inefficient cooling (e.g.\ at the disruption/capture radius, it has sufficiently slow cooling such that it cannot radiate its thermal energy in much less than the circularization time of the system). In GMCs and the dense/neutral ISM, however, the opposite is true. Observed molecular clouds and dense neutral gas commonly have thermal $\beta \ll 1$ (often $\ll 0.01$ in cold dense phases), while the turbulence is highly super-sonic and closer to trans- or modestly super-\Alf{ic} \citep{elmegreen:2004.obs.ism.turb.review,heiles.troland:2005.diffuse.ism.magnetic,mckee:2007.sf.theory.review,federrath:2012.turb.sims.sf,crutcher:2012.magnetic.fields.mc.review,beck:2015.b.field.review,krumholz.federrath:2019.review.bfields.sf}. And the cooling time for these phases, at the radii and densities of interest for the initial capture/circularization, are orders-of-magnitude shorter than the circularization times. 

This can completely change the predicted initial conditions and initial structure of accretion disks that form upon accretion/capture/disruption of said clouds \citep[see][]{pariev:2003.mag.dominated.disk.models,begelman.pringle:2007.acc.disks.strong.toroidal.fields,johansen.levin:2008.high.mdot.magnetized.disks,gaburov:2012.public.moving.mesh.code,salveson:2016.sims.mri.dominated.bfield.disks,kudoh:2020.strong.b.field.agn.acc.disk.sims.compare,huang:2023.rmhd.sims.rad.similar.mag.pressure,begelman:2023.magnetic.bh,begelman:2024.magnetized.disk,guo:2024.fluxfrozen.disks.lowmdot.ellipticals,kaaz:2024.hamr.forged.fire.zoom.to.grmhd.magnetized.disks,shi:2024.seed.to.smbh.case.study.subcluster.merging.pairing.fluxfrozen.disk,hopkins:superzoom.disk,hopkins:superzoom.analytic,hopkins:multiphase.mag.dom.disks,squire:2024.mri.shearing.box.strongly.magnetized.different.beta.states,luo:2024.magnetically.dominated.disk.like.our.zoomins.zoomin.on.first.supermassive.star.situation,wang:2025.hypermagnetized.circumbinary.disk.flux.frozen.cavity.to.pc.scales}. 
In this paper, we therefore revisit the simple analytic scalings for the initial conditions of circularized material at these capture/disruption radii, for more plausible low-$\beta$ accreting material. We consider the range from small initial BHs ($M_{\rm BH} \ll M_{\rm cl}$) through very large initial BHs ($M_{\rm BH} \gg M_{\rm cl}$), and the consequences of the fact that there must be a broad spectrum of cloud sizes and properties. We discuss the broader consequences in both an ISM and cosmological/galaxy formation context, as well as the fundamental differences in the scalings here between the accretion of cold ISM material with $\beta \ll 1$, compared to something like accretion from stars (relevant for TDEs, XRBs, CVs) with $\beta \gg 1$.

\section{Setup}
\label{sec:setup}

Table~\ref{tbl:variables} summarizes key variables used throughout.

\begin{deluxetable}{cl}
\tabletypesize{\footnotesize}
\tablecaption{Key variables and definitions.\label{tbl:variables}}
\tablehead{\colhead{Variable} & \colhead{Definition}}
\startdata
$M_{\rm BH}$ & Black hole mass \\
\hline
$M_{\rm cl}$, $R_{\rm cl}$ & Cloud mass, effective radius $\equiv (M_{\rm cl}/\pi\Sigma_{\rm cl})^{1/2}$ \\
$\Sigma_{\rm cl}$ & Cloud mean surface density \\
$v_{\rm cl}$ & Cloud velocity dispersion $\approx (G M_{\rm cl}/R_{\rm cl})^{1/2}$ \\
$v_{A,\,\rm cl}$ & Cloud rms \Alf\ speed $\approx \alpha_{\rm cl,\,B}^{1/2}\,v_{\rm cl}$ \\
$\alpha_{\rm cl,\,B}$ & Cloud magnetic-to-gravitational energy ($\sim 0.1$--$1$) \\
\hline
$\Delta v$ & BH-cloud relative velocity \\
$v_{e}$, $\tilde{v}$ & Effective encounter velocity $(v_{\rm cl}^{2} + \Delta v^{2})^{1/2} \equiv \tilde{v}\,v_{\rm cl}$ \\
$r_{t}$ & Tidal/capture/circularization radius \\
$v_{K}$ & Keplerian velocity at $r_{t}$: $(G M_{\rm BH}/r_{t})^{1/2}$ \\
$M_{\rm capt}$ & Captured (circularized) mass \\
\hline
$\Sigma_{d}$, $\rho_{d}$ & Disk surface density, midplane density \\
$H_{d}$ & Disk scale height at $r_{t}$ ($\rho_{d} = \Sigma_{d}/2 H_{d}$) \\
$Q_{d}$ & Disk Toomre parameter \\
$B_{\phi/z/R,d}$ & Disk toroidal/vertical/radial magnetic field \\
$v_{A,\,d}$,\,$\beta_{d}$ & Disk \Alf\ speed ($|B_{\phi,d}|/\sqrt{4\pi\rho_{d}}$), $\beta_{d} \equiv c_{s,d}^{2}/v_{A,d}^{2}$ \\
$\dot{M}_{d}$, $\dot{m}$ & Disk accretion rate (Eddington-normalized $\dot{M}_{d}/\dot{M}_{\rm crit}$) \\
$\Delta t_{d}$ & Disk depletion/viscous time $\equiv M_{\rm disk} / \dot{M}_{d}$ \\
\hline
$M_{0}$ & ISM Toomre/typical cloud mass $\sim \psi\,\pi\,\Sigma_{0}\,H_{0}^{2}$ \\
$\Sigma_{0}$, $H_{0}$, $\sigma_{0}$ & ISM surface density, scale height, velocity dispersion \\
$\Omega_{0}$ & ISM orbital frequency \\
$\psi$ & Cloud-to-ISM surface density ($\Sigma_{\rm cl}/\Sigma_{0} \sim 3$--$10$) \\
$\gamma_{\rm cl}$ & Cloud mass function slope (${\rm d}N/{\rm d}M \propto M^{-\gamma_{\rm cl}}$; $\sim1.5$-$2$) 
\enddata
\end{deluxetable}

Consider an encounter between a BH of mass $M_{\rm BH}$ and ``cloud'' (defined in the usual ISM sense as some self-gravitating gas overdensity) of mass $M_{\rm cl}$ with effective (spherical-equivalent) radius 
\begin{align} 
R_{\rm cl} &\equiv \sqrt{ \frac{M_{\rm cl} } {\pi \Sigma_{\rm cl}}}
\end{align}
 (in terms of the cloud-mean surface density $\Sigma_{\rm cl}$). The effective cloud turbulent+magnetic+thermal effective velocity support/scale is 
 \begin{align} 
 v_{\rm cl} = \alpha_{\rm cl}^{1/2} v_{\rm cl,\,g}  \approx \sqrt{\frac{G M_{\rm cl}}{R_{\rm cl} }} \approx v_{\rm turb} \approx v_{A,\,{\rm cl}} \,\alpha_{\rm cl,\,B}^{-1/2}
 \end{align}
 where  $v_{\rm cl,\,g}$ is the gravitational velocity of the cloud. For any reasonable/observed ISM cloud structure, turbulent support $v_{\rm turb} \approx \alpha_{\rm cl}^{1/2} v_{\rm cl,\,g} \approx v_{\rm cl,\,g}$ ($\alpha_{\rm cl}^{1/2} \sim 1$), magnetic support is defined in terms of the rms \Alf\ speed $v_{\rm A,\,{\rm cl}} \equiv \langle |{\bf B}|^{2} \rangle^{1/2} / \sqrt{4\pi \langle \rho_{\rm cl} \rangle} \equiv \alpha_{\rm cl,\,B}^{1/2} v_{\rm cl,\,g} \equiv \mathcal{M}_{A}^{-1} v_{\rm turb}$ with $\mathcal{M}_{A} \sim (\alpha_{\rm cl}/\alpha_{\rm cl,\,B})^{1/2} \sim 0.5-3$ or so \citep[references in \S~\ref{sec:intro} and][]{heiles.troland:2005.diffuse.ism.magnetic,federrath:2012.turb.sims.sf,crutcher:2012.magnetic.fields.mc.review,li:2021.gmc.bfield.review,polychronakis:2025.sf.dcf.updates.review}, and the thermal support is small ($\mathcal{M}_{\rm sonic} \equiv v_{\rm turb}/c_{s} \gg 1$, $\beta \equiv c_{s}^{2}/v_{\rm A,\,{\rm cl}}^{2} \ll 1$). 
 We define the tidal/circularization/capture radius $r_{t}$ from the BH, and denote as 
 \begin{align}
 v_{K} \equiv \Omega \, r_{t} \equiv \sqrt{\frac{G M_{\rm BH}}{r_{t}}}
 \end{align}
  the Keplerian velocity around the BH at this distance. 
 The cloud-BH mean relative velocity is 
 \begin{align}
 \Delta v &\equiv |{\bf v}_{\rm BH} - \langle {\bf v} \rangle_{\rm cl}| 
 \end{align}
  and we define 
  \begin{align}
  v_{e} &\equiv ({v_{\rm cl}^{2} + \Delta v^{2}})^{1/2} \equiv \tilde{v} v_{\rm cl}\ .
  \end{align}

Basic properties of clouds in the ISM (radiative $t_{\rm cool} \ll t_{\rm dyn}$, $\alpha_{\rm turb} \sim \alpha_{\rm cl} \sim 1$, $\mathcal{M}_{s} \gg1$, $\beta \ll 1$), mean that they must be turbulent with large density fluctuations and ``sub-clouds'' embedded down to at least the sonic scale of the turbulence. 
This is observed, with the sub-clouds following similar scale-free scalings to the parent cloud population, in the Milky Way and nearby galaxies \citep{mac-low:2004.turb.sf.review,coletta:2025.mw.clump.core.mf.relation.to.imf,traficante:2026.30dor.gmc.clouds.clumps.spectrum}, high-redshift gas-rich galaxies \citep{swinbank:clumps,livermore:2015.highz.clump.galaxies.internal.structure,zhu:2026.jades.clumps}, and is one of the more robust predictions of theory and numerical simulations \citep{hennebelle:2008.imf.presschechter,federrath:2012.turb.sims.sf,hopkins:frag.theory,guszejnov:cmf.imf,oklopcic:clumpy.highz.gals.fire.case.study.clumps.not.long.lived}. 
These ``clouds within clouds'' will play an important role in our derivation below, but unless otherwise specified we take ``cloud'' to refer to the largest parent self-gravitating structure enclosing the defined quantities above (more formally, the ``first-crossing'' structure at a given mass scale, as defined for embedded self-gravitating turbulent systems in \citealt{hopkins:excursion.ism,hopkins:excursion.imf}). For any reasonable/realistic turbulent or fluctuation or velocity/magnetic structure (see references above), most of the power in velocity $\langle |{\bf v}|^{2} \rangle^{1/2} \sim v_{\rm turb,\,\rm cl}$ or magnetic field $\langle |{\bf B}|^{2} \rangle^{1/2} \sim B_{\rm cl}$ will be on the largest scales $\sim R_{\rm cl}$. For ${\bf B}$ and ${\bf v}$, this sets the effective mean fields (up to the additive constant $\Delta {\bf v}$ in ${\bf v}$) seen by any ``sub-volume'' cloud patch interacting with the BH. 

In the different scenarios considered, the BH will capture and circularize a mass $M_{\rm capt}(t)$ at the radius $r_{t}$ at some time $t$ after the encounter begins, with some vertical effective velocity support $v_{d}^{2} \equiv v_{\rm turb,\,d}^{2} + v_{A,\,d}^{2} + c_{s,\,d}^{2}$. 
This will lead to a ``disk'' with some effective surface density $\Sigma_{d}$, Toomre $Q_{d}$ parameter, scale height $H_{d}$
\begin{align}
\Sigma_{d} &\sim \frac{M_{\rm capt}(t)}{2\pi r_{t} \Delta R} \ , \\
Q_{d} &\sim \frac{v_{d} \Omega_{d} } {\pi G \Sigma_{d}}  \sim \frac{H_{d}}{r_{t}} \frac{2 \Delta R}{r_{t}} \, \frac{M_{\rm BH}}{M_{\rm capt}[t]} \ , \\ 
H_{d} &\sim \frac{v_{d}}{ v_{K}} \,r_{t} \ .
\end{align}
$\Delta R \sim r_{0} \approx {\rm MIN}[r_{t},\,R_{\rm cl}]$ is the width of the initial circularizing structure in the $R$ direction (comparable to the capture or cloud radius, whichever is smaller, as the captured stream circularizes and spreads radially).

Being radiative $c_{s,\,d} \approx c_{s,\,\rm cl} \ll v_{d},\,v_{\rm cl}$ will be small. The change in turbulent velocities is somewhat more difficult to address but since the turbulent energy is the dominant kinetic energy and the capture process is rapid (occurring on $\mathcal{O}(1)$ dynamical times, which necessarily means $\mathcal{O}(1)$ largest-eddy turnover times for the scalings below), energy conservation would imply approximately ``isothermal'' turbulence as the initial condition, $v_{\rm turb,\,d} \sim v_{\rm turb,\,\rm cl}$. More precisely, because the capture/circularization involves differential gravitational acceleration across the cloud and shearing into a torus, the initial turbulence will also be augmented by any ``stretching'' of the cloud in the $\phi$ direction, but this tends to drive $v_{\phi}$ modes rather than the $v_{z}$ relevant for $H_{d}$, so our approximation is reasonable to leading order.

In the initial capture/disruption, that rapidity implies that the dominant process influencing ${\bf B}$ should be flux-freezing. Assuming there is no global strongly-preferred direction of ${\bf B}$ in the cloud (as observed; \citealt{crutcher:2012.magnetic.fields.mc.review}), this implies 
\begin{align}
B_{\phi,\,d} &\sim B_{\phi,\,0} \frac{ \ell_{z,\,0} \ell_{R,\,0}}{\ell_{z,\,d} \ell_{R,\,d}} \sim   \frac{B_{\rm cl}\,r_{0}^{2} }{ H_{d} \Delta R} \sim {B_{\rm cl} } \frac{r_{0}}{H_{d}} \frac{r_{0}}{\Delta R}   \ , \\
B_{z,\,d} &\sim B_{z,\,0} \frac{\ell_{\phi,\,0} \ell_{R,\,0} }{ \ell_{\phi,\,d} \ell_{R,\,d} }\sim   \frac{B_{\rm cl}\, r_{0}^{2} }{ 2\pi r_{t} \Delta R } \sim \frac{B_{\rm cl} }{2\pi} \frac{r_{0}}{r_{t}}  \frac{r_{0}}{\Delta R}  \ , \\
B_{R,\,d} &\sim B_{R,\,0} \frac{\ell_{z,\,0} \ell_{\phi,\,0} }{\ell_{z,\,d} \ell_{\phi,\,d} } \sim   \frac{B_{\rm cl}\,r_{0}^{2} }{ 2\pi r_{t} H_{d} } \sim  \frac{B_{\rm cl}}{2\pi} \frac{r_{0} }{ r_{t}} \frac{r_{0}}{ H_{d} } \ , 
\end{align}
where $r_{0}$ is the size of the cloud patch from which the largest coherent fields are captured. Since $H_{d} \le r_{0} \le \Delta R \le r_{t}$, this means that for isotropic initial cloud fields one will always have ``initial'' disk fields with 
\begin{align}
B_{z,\,d} \lesssim B_{R,\,d} \lesssim B_{\phi,\,d}\ .
\end{align}
This plus the disk density $\rho_{d} \equiv \Sigma_{d} / 2 H_{d}$ determines the disk \Alf\ speed support in the vertical direction, $v_{A,\,d} \equiv v_{A,\,\phi,\,d} \equiv B_{\phi,\,d} / \sqrt{4\pi \rho_{d}}$.

Given that we'll argue dominant stress/pressure is turbulent+in-plane magnetic, we will take the effective transport speed to be order the dominant support speed, i.e.\ $\nu_{\rm eff} \sim v_{d}^{2}/\Omega$ up to an order-unity efficiency factor. With this convention we can crudely estimate the initial disk accretion rate $\dot{M}_{d}$ and depletion time $\Delta t_{d}$ as 
\begin{align}
\label{eqn:mdot.general} \dot{M}_{d} &\sim 3\pi \nu_{\rm eff} \Sigma_{d} \sim \left( \frac{H_{d}}{ r_{t}} \right)^{2} M_{\rm capt}\, \Omega \ , \\
\Delta t_{d} &\sim \frac{M_{\rm capt}}{\dot{M}_{d}} \sim \Omega^{-1} \left( \frac{H_{d}}{ r_{t}} \right)^{-2} \ .
\end{align}
We can also define this relative to the critical accretion $\dot{M}_{\rm crit} \equiv M_{\rm BH} / t_{S}$ for $t_{S} \sim 5 \times 10^{7}\,{\rm yr}$ (standard for a BH with radiative efficiency $\epsilon_{r} \sim 0.1$): 
\begin{align}
\dot{m} &\equiv \frac{\dot{M}_{d}}{\dot{M}_{\rm crit}} = \frac{\dot{M}_{d} t_{S}}{M_{\rm BH}} \ . 
\end{align}

We will consider a mass spectrum of clouds motivated by observations \citep[references above and][]{solomon:gmc.scalings,rosolowsky:2005.gmc.mass.function,bolatto:2008.gmc.properties,colombo:2014.gmc.catalog,miville:2017.gmc.mf,leroy:2025.cloud.sfe.compilation.phangs} and generic theoretical predictions for scale-free systems \citep{hopkins:excursion.ism,hopkins:frag.theory,krumholz.mckee:2005.turb.sf.theory},
\begin{align}
\frac{d N_{\rm cl} }{ d M_{\rm cl}} \equiv (2-\gamma_{\rm cl})\,\frac{\rho_{0}}{M_{0}^{2}}\,\left( \frac{M_{\rm cl}}{M_{0}} \right)^{-\gamma_{\rm cl}}
\end{align}
 with $1.5 < \gamma_{\rm cl} < 2$ and $M_{\rm min} < M_{\rm cl} < M_{0}$ ($M_{\rm min} \ll M_{0}$ is irrelevant as long as it is small), and $\rho_{0} \equiv \Sigma_{0} / H_{0}$ is the total ISM-integrated mass density of clouds, $\Sigma_{0}$ the ISM gas surface density, $H_{0}$ is scale-height $\equiv \sigma_{0} / \Omega_{0}$ and the ISM Toomre 
 \begin{align}
 Q_{0} \sim 1\approx \frac{\sigma_{0} \Omega_{0} } {\pi G \Sigma_{0} } \sim \frac{ \Omega_{0}^{2} }{ 2\pi G \rho_{0}}.
 \end{align} 
 Dynamical equilibrium means the typical 
 \begin{align}
 \Delta v \sim \sigma_{0} \sim H_{0}\, \Omega_{0}.\ 
 \end{align}
  Another extremely general expectation and observation \citep{larson:gmc.scalings,solomon:gmc.scalings,bolatto:2008.gmc.properties,heyer:2009.gmc.trends.w.surface.density} is a linewidth-size relation akin of the form $v_{\rm cl} \propto R_{\rm cl}^{1/2}$ (arising from any weakly-correlated fluctuations), all of which leads to the scalings: 
  \begin{align}
  M_{0} &\sim \psi \pi \Sigma_{0} H_{0}^{2}\ , \\ 
  \Sigma_{\rm cl} &\sim \psi \Sigma_{0}\ , \\
  R_{\rm cl} &\sim H_{0}\,(M_{\rm cl}/M_{0})^{1/2} \  , \\ 
  v_{\rm cl} &\sim \sigma_{0}\,(\psi/Q_{0})^{1/2} \,(M_{\rm cl}/M_{0})^{1/4} 
  \end{align}
   with $\psi \sim 3-10$ depending on details \citep{hopkins:excursion.ism,hopkins:2013.turb.sf.mol.hi,hopkins:frag.theory,cafg:sf.fb.reg.kslaw}.  

The effective cross-section $\sigma_{\rm BH-cl}$ for an some strong BH-cloud ``encounter'' is given by the maximum of either the ``capture'' radius or cloud size, 
\begin{align}
\sigma_{\rm BH-cl} &\sim \pi\, {\rm MAX}[R_{\rm cl},\,r_{t}]^{2}\ .
\end{align}

Given the cloud population, the rate of such ``encounter'' events, per unit cloud mass, can be estimated as 
\begin{align}
\frac{d\dot{N}_{\rm events}}{d\ln{M_{\rm cl}}} &\sim \sigma_{\rm BH-cl}\, \Delta v\, \frac{dN_{\rm cl}}{d\ln{M_{\rm cl}} }\ , 
\end{align}
and if each event has a ``duration'' $\Delta t_{\rm event}$, the duty cycle scales as
\begin{align}
\frac{d f_{\rm events}}{d\ln{M_{\rm cl}} }& \sim \Delta t_{\rm event} \dot{N}_{\rm events} \ , 
\end{align}
so the long time-and-event averaged mass capture rate scales like
\begin{align}
\frac{d\dot{M}_{\rm events}}{d\ln{M_{\rm cl}}} &\sim \dot{M}_{\rm capt} \Delta t_{\rm event} \dot{N}_{\rm events} \ .
\end{align}
Each of these can be integrated over the cloud mass spectrum to yield a total, e.g. 
\begin{align}
\dot{M}_{\rm events} \equiv \int \frac{d\dot{M}_{\rm events}}{d\ln{M_{\rm cl}}} d\ln{M_{\rm cl}}\ .
\end{align}

\begin{figure*}
\centering
\includegraphics[width=0.95\textwidth]{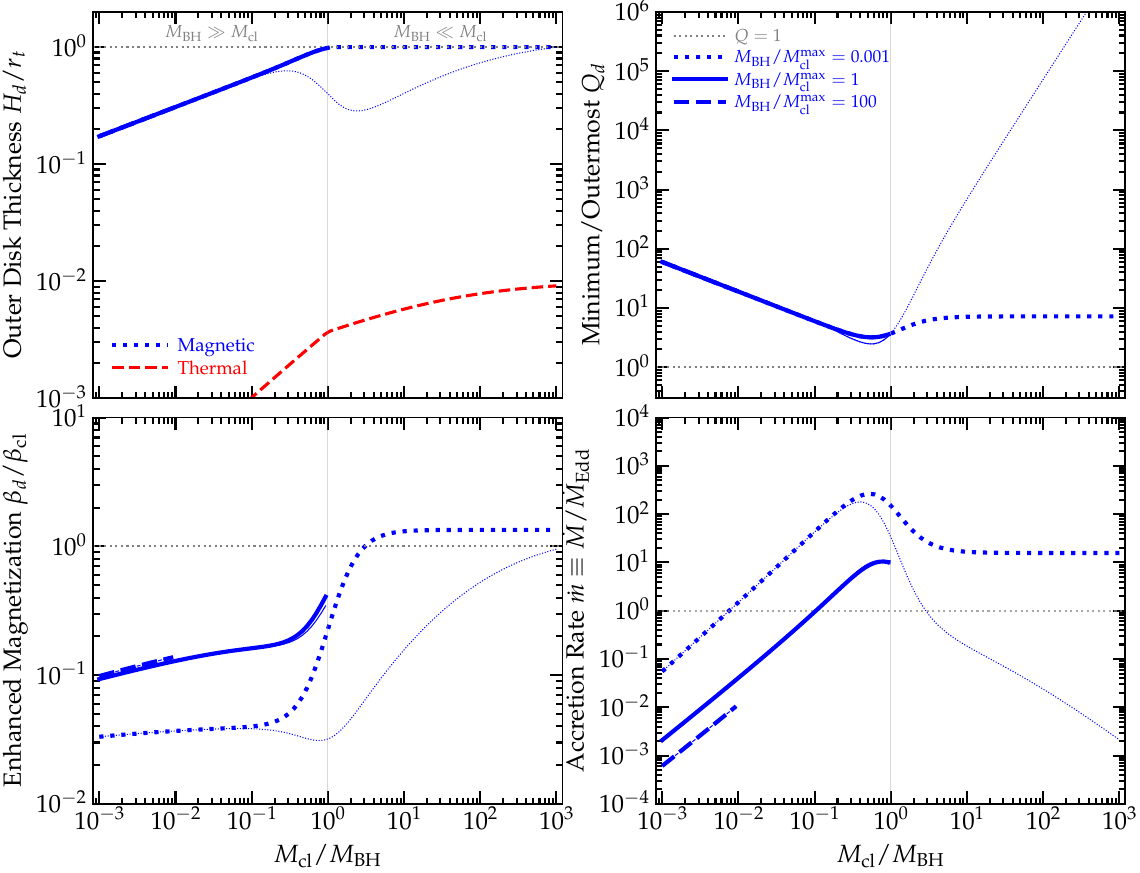}
\caption{Predicted initial disk properties at the circularization radius as a function of cloud-to-BH mass ratio $M_{\rm cl}/M_{\rm BH}$, for fiducial parameters ($\alpha_{\rm cl,\,B}=0.3$, $\psi=5$, $Q_{0}=1$, and different BH-to-ISM maximum cloud mass $M_{\rm cl}^{\rm max}$ ratios labeled). 
{\em Top left:} Scale height/aspect ratio $H_{d}/r_{t}$ (\textit{blue}). A crude estimate of the thermal value is compared (\textit{red}). 
We also contrast the prediction (differing for $M_{\rm BH} \lesssim M_{\rm cl}$) from the lower-accretion-rate, higher-duty-cycle ``turbulent Bondi-Hoyle'' mode (\textit{thin}) vs.\ the ``leap-frog'' rapid-growth mode encountering sub-clouds of mass $\sim M_{\rm BH}$ (\textit{thick}). 
{\em Top right:} Toomre $Q_{d}$ parameter at circularization (including magnetic support). The disk is always (initially) gravitationally stable ($Q_{d} \gg 1$). 
{\em Bottom left:} Ratio of disk-to-progenitor cloud plasma $\beta = P_{\rm thermal} / P_{\rm mag} \sim c_{s}^{2} / v_{A}^{2}$. In all regimes $\beta_{d} \lesssim \beta_{\rm cl} \ll 1$ -- the disk at circularization is as-or-more magnetically dominated as the progenitor cloud. 
{\em Bottom right:} Eddington-normalized accretion rate $\dot{m}$. High-$\dot{m}$ accretion is generic to broadly comparable-mass encounters (in the ``leapfrog'' mode, $M_{\rm cl} \gg M_{\rm BH}$ but sub-cloud $M_{\rm cl}^{\rm sub} \sim M_{\rm BH}$). 
\label{fig:disk.properties}}
\end{figure*}

\begin{figure}
\centering
\includegraphics[width=\columnwidth]{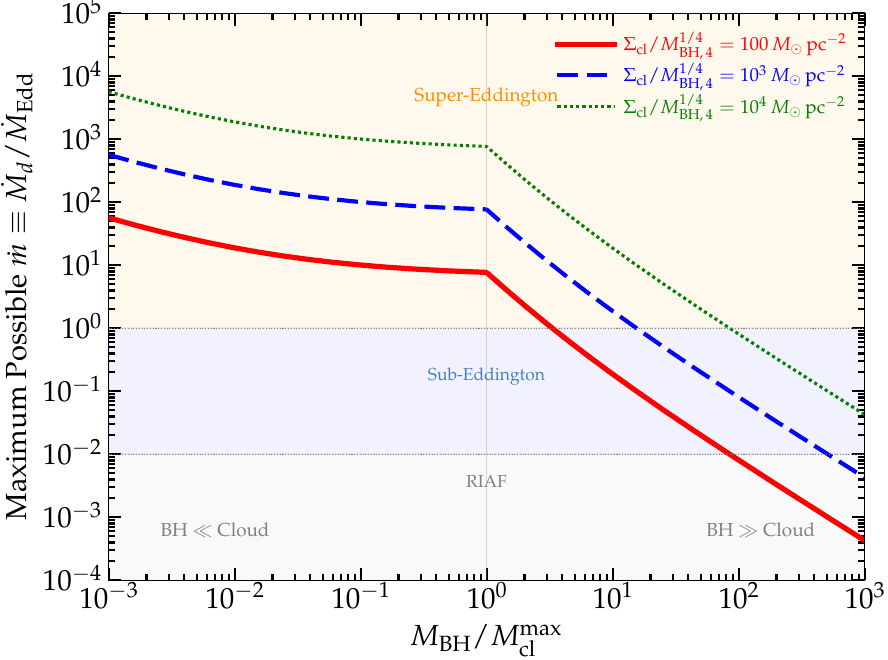}
\caption{Peak predicted Eddington-normalized accretion rate $\dot{m}$ (defined in terms of $\dot{M}_{d}$ from the outer disk, marginalizing over encounters with all clouds of mass $0 < M_{\rm cl} < M_{\rm cl}^{\rm max}$ and the leapfrog-vs.-turbulent Bondi mode), as a function of the BH-to-ISM maximum cloud mass $M_{\rm cl}^{\rm max}$ ratio.
We compare different cloud surface densities (assuming all clouds have the same $\Sigma_{\rm cl}$)  normalized by BH mass $\Sigma_{\rm cl} / M_{\rm BH,\,4}^{1/4}$ ($M_{\rm BH,\,4} \equiv M_{\rm BH} / 10^{4}\,M_{\odot}$). 
In the regime of massive BHs and quenched galaxies (small $M_{\rm cl}^{\rm max}$, $\Sigma_{\rm cl}$), $\dot{m}$ from cloud-capture is always low. 
Gas-rich, dense galaxies with high-$M_{\rm cl}^{\rm max}$ and $\Sigma_{\rm cl}$ (e.g.\ high-redshift gas-rich massive galaxies, starburst nuclei) can produce supercritical $\dot{m} \gg 1$ accretion in the outer disk.
\label{fig:regime.diagram}}
\end{figure}

We will consider three regimes of the ratio of BH to cloud mass: $M_{\rm BH} \ll M_{\rm cl}$, $M_{\rm BH} \sim M_{\rm cl}$, and $M_{\rm BH} \gg M_{\rm cl}$, as these allow us to make different approximations and give rise to different behaviors in some important respects. Fig.~\ref{fig:disk.properties} summarizes the resulting disk properties across all regimes, and Fig.~\ref{fig:regime.diagram} shows the predicted peak Eddington-normalized accretion rate as a function of the BH-to-cloud mass ratio.

\section{Low-Mass BHs: $M_{\rm BH} \ll M_{\rm cl}$}
\label{sec:lowmass.bhs}

Consider the smallest BH-to-cloud mass ratios $M_{\rm BH} \lesssim 0.01\,M_{\rm cl}$. This is relevant for stellar \&\ LIGO-mass seed BHs, as well as IMBHs encountering the most massive cloud complexes in galaxies (especially for giant GMC-like ``clump'' complexes in high-redshift galaxies, and central molecular reservoirs in starburst galaxies). It is convenient to work in the cloud rest frame, and consider the BH as a passerby. 

In this limit the ``tidal radius'' not directly relevant for accretion, as $r_{\rm tidal} \sim (M_{\rm cl}/M_{\rm BH})^{1/3} R_{\rm cl}$ would imply $v_{K} \sim v_{\rm cl,\,g} (M_{\rm BH}/M_{\rm cl})^{1/3} \ll v_{\rm cl,\,g}$, and cloud material with $\alpha_{\rm cl} \sim 1$ would be unbound. Instead the relevant ``capture'' and ``circularization'' radii are order of magnitude given by those interior to which material can be bound to the BH (and/or has sufficiently low angular momentum to circularize within that orbit),  i.e.\ the radius where 
\begin{align}
v_{K} \sim v_{e}\ ,
\end{align}
which is just the Bondi-Hoyle-like radius 
\begin{align}
r_{t} \sim \frac{G M_{\rm BH}}{v_{e}^{2}} \sim \frac{M_{\rm BH}}{M_{\rm cl} \tilde{v}^{2}}\,R_{\rm cl}.
\end{align}
Because $r_{t} \ll R_{\rm cl}$, we have
\begin{align}
\sigma_{\rm BH-cl} \sim \pi R_{\rm cl}^{2}.
\end{align}
The BH captures gas from a cylinder of long-axis length $\sim R_{\rm cl}$ and cylindrical circular radius $r_{t}$. Material captured on either ``side'' of the BH will have relative velocities $\pm \Delta v$ equal-and-opposite, which would cancel if the cloud were homogeneous, but because the cloud has both inhomogeneous density and velocity structure at the $\mathcal{O}(1)$ level on all size-scales of interest (except for vanishingly small BH masses relative to the cloud), the ``net'' specific velocity and therefore angular momentum of the captured material will be within a factor of a few of $v_{e}$ and $\sim r_{0} v_{e} \sim r_{t} v_{K}$. The circularization radius will therefore scale with the capture radius (a factor of a few smaller), $r_{\rm circ} \sim \zeta r_{t}$. 

Because $r_{t} \ll R_{\rm cl}$, the captured ${\bf B}$ and ${\bf v}$ will be coherent to leading-order on scales $\mathcal{O}(r_{t})$. The captured mass will initially grow at a rate $\dot{M}_{\rm capt} \sim \langle \rho_{\rm cl} \rangle  \pi r_{t}^{2} v_{e}$. 
At $t\rightarrow 0$ this means $\dot{M}_{d} \ll \dot{M}_{\rm capt}$, so mass ``piles up'' in the disk until $\dot{M}_{d} \approx \dot{M}_{\rm capt}$, implying a steady-state $M_{\rm capt} \sim \dot{M}_{\rm capt} r_{t}^{2} /(v_{e} H_{d}) \sim \langle \rho_{\rm cl} \rangle  \pi r_{t}^{5}/H_{d}^{2}$. This implies $\rho_{d} \sim \Sigma_{d}/(2 H_{d}) \sim \langle \rho_{\rm cl} \rangle / (4 (H_{d}/r_{t})^{3})$, which with $B_{\phi,\,d} \sim B_{\rm cl}/(H_{d}/r_{t})$ implies the \Alf\ speed giving vertical support is $v_{A,\,\phi,\,d} \sim 2 (H_{d}/r_{t})^{1/2} v_{A,\,{\rm cl}}$, so $v_{A,\,\phi,\,d} /v_{K} \sim 2 (H_{d}/r_{t})^{1/2} \tilde{v}^{-1}$. This is $\mathcal{O}(1)$, so compared to the initial thermal support of the disk $c_{s,\,d}/v_{K} \sim c_{s,\,\rm cl}\,\tilde{v}^{-1} \ll 1$ shows the latter is negligible. The turbulent support requires we split $v_{\rm turb,\,\rm cl}$ into the mean motion term $\sim v_{\rm cl}$ coming from modes on scales $R_{\rm cl} \gg r_{t}$, which appears in $v_{e}$ added in quadrature to $\Delta v$ as a bulk term, from the actual dispersion on the capture scale, $v_{\rm turb,\,d} \sim v_{\rm turb}(r_{t}) \sim v_{\rm cl}\,(r_{t}/R_{\rm cl})^{1/2} \sim v_{\rm cl}\,(M_{\rm bh}/M_{\rm cl})^{1/2}\,\tilde{v}^{-1}$. So long as $\alpha_{\rm cl,\,B}$ (expected to be $\sim 0.05-1$) is not very small $\ll 0.25\,(M_{\rm BH}/M_{\rm cl})^{1/2}$, which should be true, then this implies the initial magnetic support is more important for determining the initial $H_{d} \sim v_{A,\,\phi,\,d}/\Omega$. Assuming this, we obtain
\begin{align}
\frac{H_{d}}{r_{t}} &\sim \frac{4 \alpha_{\rm cl,\,B}}{\tilde{v}^{2}} \sim \mathcal{O}(1) \ , \\
B_{\phi,\,d} &\sim  \frac{\tilde{v}^{2}}{4 \alpha_{\rm cl,\,B}}  B_{\rm cl} > B_{\rm cl} \ , \\
B_{z,\,d} &\sim   \frac{1}{2\pi}  B_{\rm cl} < B_{\rm cl}\ , \\
B_{R,\,d} &\sim \frac{\tilde{v}^{2}}{2\pi \times 4 \alpha_{\rm cl,\,B}} B_{\rm cl}  \ , \\
M_{\rm capt} &\sim M_{\rm BH} \,  \left( \frac{M_{\rm BH}}{8\,\alpha_{\rm cl,\,B} \tilde{v} M_{\rm cl}} \right)^{2} \ , \\ 
Q_{d} &\sim \frac{256\pi}{3}\,\alpha_{\rm cl,\,B}^{3} \left( \frac{M_{\rm cl}}{M_{\rm BH}}\right)^{2} \gg 1 \ , \\
\Delta t_{d} &\sim \Omega^{-1} \left( \frac{\tilde{v}^{2}}{4 \alpha_{\rm cl,\,B}} \right)^{2} \ , \\ 
\dot{M}_{d} &\sim M_{\rm BH} \frac{1}{4\tilde{v}^{3}} \frac{M_{\rm BH}}{M_{\rm cl}} \frac{v_{\rm cl}}{R_{\rm cl}} \ . 
\end{align}

Note that we can confirm that the time to initially ``sweep up'' this much mass and/or replenish the disk, $\sim M_{\rm capt} / \dot{M}_{\rm capt} \sim \Delta t_{d} \sim (r_{t}/v_{e})\,(\tilde{v}^{4}/16\alpha_{\rm cl,\,B}^{2})$ above is much smaller than the cloud crossing time $\sim R_{\rm cl}/\Delta v$, so this quasi-steady capture+depletion is a reasonable assumption. 
This means the encounter duration will be $\Delta t_{\rm event} \sim R_{\rm cl}/\Delta v$, which allows us to estimate: 
\begin{align}
\frac{d\dot{N}_{\rm events}}{d\ln{M_{\rm cl}}} &\sim  \pi R_{\rm cl}^{2}  \Delta v\, \frac{dN_{\rm cl}}{d\ln{M_{\rm cl}} } \sim \frac{(2-\gamma_{\rm cl})}{\psi} \Omega_{0} \left( \frac{M_{\rm cl}}{M_{0}} \right)^{2-\gamma_{\rm cl}} \ , \\ 
\frac{d f_{\rm events}}{d\ln{M_{\rm cl}} }& \sim \frac{(2-\gamma_{\rm cl})}{\psi} \left( \frac{M_{\rm cl}}{M_{0}} \right)^{5/2-\gamma_{\rm cl}}
 \ , \\
\frac{d\dot{M}_{\rm events}}{d\ln{M_{\rm cl}}} &\sim 
 \frac{1}{\tilde{v}^{3}} \frac{M_{\rm BH}^{2}}{M_{0}} \Omega_{0} \,
\frac{(2-\gamma_{\rm cl})}{4 (\psi Q_{0})^{1/2}} \left( \frac{M_{\rm cl}}{M_{0}} \right)^{5/4-\gamma_{\rm cl}}
 \ .
\end{align}
For reasonable numbers, this implies $\dot{N}_{\rm events}$ and the duty cycle are driven by the most massive (more volume-filling) complexes, while $\dot{M}_{\rm events}$ is dominated by low-mass clouds (owing to the strong inverse dependence of $r_{t}$ on $M_{\rm cl}$) until it is suppressed at very low $M_{\rm cl}$ because of the $\tilde{v}$ dependence. For reasonable numbers the peak contribution to $\dot{M}_{\rm event}$ would come from $M_{\rm cl} \sim 0.03\,M_{0}$, with peak amplitude $\dot{M}_{\rm events} \sim 0.04\, M_{\rm BH}\,(M_{\rm BH}/M_{0})\,\Omega_{0}$.

\subsection{Super-Virial Velocities, or Low-Surface Density Clouds: Turbulent Bondi-Hoyle Accretion}
\label{sec:lowmass.bhs:lowacc}

Given these scalings, for $\Delta v \gg v_{e}$, the problem can be reduced to a BH moving on an unperturbed trajectory through the ``infinite'' cloud medium for a time $\Delta t \sim R_{\rm cl} / \Delta v$, i.e.\ the turbulent Bondi-Hoyle problem.  

This limit is realized when $\Delta v \gg v_{\rm cl}$, i.e.\ the BH moves through the cloud at highly super-virial velocities. This occurs naturally for the smallest clouds in the mass spectrum, where $v_{\rm cl} \propto M_{\rm cl}^{1/4}$ becomes small relative to the ambient velocity dispersion $\Delta v \sim \sigma_{0}$. It also applies to BHs with large peculiar velocities (e.g.\ BHs that have been gravitationally scattered to super-virial velocities, or BHs in the outer regions of galactic disks where $\Delta v \gg v_{\rm cl}$ for local clouds).

This also implies (in this limit) 
\begin{align}
\dot{m} \sim \frac{2}{\tilde{v}^{3}} \left( \frac{M_{\rm BH}}{M_{\rm cl}} \right) \left( \frac{M_{0}}{M_{\rm cl}} \right)^{1/4} \frac{\Omega_{0}}{(10^{7} \, {\rm yr})^{-1}} \ll 1 \ .
\end{align}
In this limit, let us compare the accretion/depletion timescale to the timescales for Parker flux-loss from the midplane and/or turbulent dissipation. For a $\beta \ll 1$, toroidally-magnetized, isothermal ideal-MHD disk with no other source of turbulence or support (not necessarily valid in reality of course), both local \citep{squire:2024.mri.shearing.box.strongly.magnetized.different.beta.states} and global \citep{guo:2024.fluxfrozen.disks.lowmdot.ellipticals,guo:2025.toroidal.disk} experiments (as well as follow-up with other codes and numerical methods; \citealt{tomar:2024.parker.instability.acc.disks}) show that while the analytic linear growth timescale of the Parker instability is $\sim \Omega^{-1}$, the time required for the instability to actually appreciably remove mean-toroidal flux and magnetic support from the midplane is much larger, $\Delta t_{\rm Parker} \gtrsim 80\,t_{\rm orbit} \sim 500\,\Omega^{-1}$. This flux loss can be completely avoided if, for example, there are other drivers of trans-\Alf{ic} turbulence in the disk \citep{squire:2024.mri.shearing.box.strongly.magnetized.different.beta.states} or significant warps or eccentric modes \citep{kaaz:2022.grmhd.sims.misaligned.acc.disks.spin,kaaz:2024.hamr.forged.fire.zoom.to.grmhd.magnetized.disks}, or additional sources of pressure (e.g.\ radiation;  \citealt{hopkins:superzoom.disk}), but it nonetheless represents a useful ``worst case'' scenario for vertical flux transport. Even absent any dynamo action, if the accretion/inflow time $\Delta t_{d} \lesssim t_{\rm Parker}$, then flux would be replenished (and boosted via simple flux-freezing, again) faster than it can be lost. So it is useful to consider the ratio:
\begin{align}
\frac{\Delta t_{d}}{\Delta t_{\rm Parker}} &\sim \frac{\Delta t_{d} }{500\,\Omega^{-1}} \sim 10^{-4} \frac{\tilde{v}^{4}}{\alpha_{\rm cl,\,B}^{2}}\ .
\end{align}
which should be $\ll 1$ unless $\tilde{v} \gtrsim 3-10$, which in turn requires $M_{\rm cl} \ll 0.01\,(Q_{0}/\psi)^{2}\,M_{0} \sim 10^{-4} M_{0}$ for a typical $\psi \sim 10$. But from the scalings above, this means $\Delta t_{d} \gtrsim \Delta t_{\rm Parker}$ corresponds to accretion rates $\dot{m} \lesssim 2\times10^{-4}\,(\psi/Q_{0})^{1/2}\,\alpha_{\rm cl,\,B}^{-1}\,\Omega_{0}/(10^{7}\,{\rm yr})^{-1}$, orders-of-magnitude below Eddington and well into the regime where we would expect the accretion flow to become radiatively inefficient and spherical (so $\Delta t_{d}$ and the Parker instability itself do not really apply anymore). Moreover, recalling that we already imposed $M_{\rm BH} \lesssim 0.01\,M_{\rm cl}$, this requires something like $M_{\rm BH} \lesssim 10^{-6}\,M_{0}$ ($M_{0}$ the characteristic cloud mass), so would only be relevant for extremely low-mass BHs approaching stellar-mass, experiencing Bondi-like accretion, at extremely sub-critical rates.

Another interesting timescale is the turbulent flux transport time. Assuming only flux-freezing, the timescale for $B_{\phi}$ to grow and/or new flux to be advected in the midplane is $t^{-1} \sim \langle v_{R} \rangle /R \sim \Delta t_{d}^{-1}$. In principle, turbulence can transport flux or give rise to an effective turbulent resistivity which invalidates the assumption of flux-freezing. As shown in the studies cited above, this does not actually necessarily lead to weaker fields, and can in fact support dynamo action. But let us take the worst-case scenario as an upper limit, and assume a turbulent resistivity $\partial_{t} B \sim -\eta_{\rm turb} k_{B_{\phi}}^{2} B$ with $k_{B_{\phi}} \sim  |\partial_{R} B_{\phi} / B_{\phi}  | \sim 1 /\ell_{R}$ and $\eta_{\rm turb} \sim \alpha_{t,\,d}^{\prime} v_{\rm turb,\,d}^{2} / \Omega$ (and again to be ``worst-case,'' take $\alpha_{t,\,d}^{\prime} \sim 1$). This gives a radial turbulent flux-transport or decay timescale of $\Delta t_{\rm turb,\,res} \sim (\ell_{R} \Omega/v_{\rm turb,\,d})^{2} \Omega^{-1}$. Comparing this to the accretion timescale gives: 
\begin{align}
\frac{\Delta t_{d}}{\Delta t_{\rm turb,\,res}} &\sim \frac{(M_{\rm BH}/M_{\rm cl})}{(4\alpha_{\rm cl,\,B})^{2}}
\end{align}
which will be $\lesssim 1$ for the BH masses and $\alpha_{\rm cl,\,B}$ of interest here (only $\gg 1$ for extremely small $\alpha_{\rm cl,\,B}$, at which point the other scalings above used to derive this will not apply).


\subsection{``Lucky'' BHs in Bounded Encounters: Leap-frogging via Sub-Clouds}
\label{sec:lowmass.bhs:hiacc}

If $M_{\rm BH}$ is small compared to the cloud $M_{\rm cl}$, then the captured mass according to turbulent Bondi-Hoyle-like accretion becomes even smaller, $M_{\rm capt} \sim M_{\rm BH}\,(M_{\rm BH}/M_{\rm cl}) \rightarrow 0$, and as shown above, the accretion rates become vanishingly small. But in such massive clouds, there are always a large number of ``sub-clumps'' or sub-clouds with masses up to the parent cloud mass $M_{\rm cl}$. As shown in numerical simulations of BH accretion in turbulent, self-gravitating clouds with ``live'' star formation and stellar feedback \citep[e.g.][and references therein]{shi:2022.hyper.eddington.no.bhfb,shi:2024.seed.to.smbh.case.study.subcluster.merging.pairing.fluxfrozen.disk,shi:2024.imbh.growth.feedback.survey}, this means that statistically, at some point (for small enough $M_{\rm BH}/M_{\rm cl}$ initially), the accretion becomes dominated by encounters with more-massive sub-clouds, with masses more comparable to $M_{\rm BH}$, because while these encounters are less common, they enable so much more efficient BH capture and growth. We justify this and develop a simple model for this regime below.

Consider encounters between the BH and ``sub-clouds'' (clouds, clumps, cores, etc.) within the parent cloud complex. This is analogous to the case we have considered above, so we will only briefly outline the scalings. Subclouds form a mass function: 
\begin{align}
\frac{d n_{\rm sub} }{ d M_{\rm sub}} \equiv (2-\gamma_{\rm sub})\,\frac{f_{\rm sub}\langle \rho_{\rm cl} \rangle}{M_{\rm cl}^{2}}\,\left( \frac{M_{\rm sub}}{M_{\rm cl}} \right)^{-\gamma_{\rm sub}}
\end{align}
with $f_{\rm sub} \sim 1$ and $\gamma_{\rm sub} \sim \gamma_{\rm cl}$ observed and predicted as clouds are close to self-similar over the mass range of interest (references in \S~\ref{sec:intro}). 
The number of subclouds encountered per cloud-crossing time $\Delta t_{\rm cross} \sim R_{\rm cl} / v_{e}$ is $d N/d \ln{M}_{\rm sub} \sim  \pi r_{\rm cyl}^{2} v_{e} \Delta t_{\rm cross} dn_{\rm sub}/d \ln{M}_{\rm sub}$, where the capture/interaction radius is $r_{\rm cyl} \sim {\rm MIN}[G M_{\rm BH} / \delta v_{\rm BH-sub}^{2} \ , \ R_{\rm sub}]$ in terms of the relative BH-subcloud velocity $\delta v_{\rm BH-sub}$. Since we are interested in $M_{\rm sub} \sim M_{\rm BH} \ll M_{\rm cl}$, for ``typical'' $\delta v_{\rm BH-sub} \sim v_{e} \sim v_{\rm cl} \gg \sqrt{G M_{\rm BH}/R_{\rm sub}[M_{\rm sub}=M_{\rm BH}]}$, the large $\delta v_{\rm BH-sub}$ will strongly suppress the probability of strong interaction or, equivalently, the radius $r_{\rm cyl} \ll R_{\rm sub}$ and mass accreted (in the usual Bondi-like fashion above). But in a turbulent medium there is a broad distribution of local ${\bf v}$ and so some interactions can have small $\delta v_{\rm BH-sub} \lesssim \delta v_{\rm crit} \equiv \sqrt{G M_{\rm BH}/R_{\rm sub}[M_{\rm sub}=M_{\rm BH}]}$. Assuming a Maxwellian turbulent velocity distribution in the cloud with $M_{\rm BH} \ll M_{\rm cl}$ and $\delta v_{\rm crit} \ll \langle \delta v_{\rm BH-sub}^{2} \rangle^{1/2} \sim v_{\rm cl}$, the fraction of ``slow'' encounters with $\delta v_{\rm BH-sub} \lesssim \delta v_{\rm crit}$ is approximately given by 
\begin{align} 
f_{\rm slow} 
 \sim \sqrt{\frac{2}{\pi}} \left( \frac{\delta v_{\rm crit} }{ v_{\rm cl} } \right)^{3}\ . 
\end{align}
Together this gives a probability of a ``slow'' capture encounter (effective $\tilde{v} \sim 1$) with a subcloud of mass $M_{\rm sub} \sim M_{\rm BH}$ of $P_{\rm crit} \sim f_{\rm slow} dN/d\ln{M}_{\rm sub}[M_{\rm sub} \sim M_{\rm BH}]$ of: 
\begin{align}
P_{\rm crit} &\sim \frac{3\,(2-\gamma_{\rm cl})}{2 \sqrt{2\pi}} \, \frac{\Delta t}{t_{\rm cross}} \, \left( \frac{M_{\rm BH}}{M_{\rm cl}} \right)^{7/4-\gamma_{\rm cl}} \, \left( \frac{\Sigma_{\rm cl}}{\Sigma_{\rm sub}} \right)^{1/4} \\ 
\nonumber &\sim 0.15\,\left( \frac{\Sigma_{\rm cl}}{\Sigma_{\rm sub}} \right)^{1/4}\,\frac{\Delta t}{t_{\rm cross}}
\end{align}



The probability of such a ``slow'' encounter is a weak function of most parameters, of order $\sim 1-15\%$ per cloud crossing, depending on the ratio $\Sigma_{\rm cl}/\Sigma_{\rm sub}$ and the mass function slope. The resulting growth is therefore hierarchical: the BH encounters one sub-cloud of comparable mass, grows by $\mathcal{O}(1)$, then encounters the next, and so on. One cannot assume that successive ``leaps'' are independent, because both observations and theory predict very strong auto-correlation of both density and velocity fluctuations in turbulent clouds \citep{hopkins:excursion.clustering,hopkins:frag.theory}. This is inevitable in any self-gravitating or scale-free cascade. Simulations show this hierarchical growth in the context of star cluster fragmentation/mergers and seed BH growth \citep{shi:2021.imbh.mass.budget,shi:2024.seed.to.smbh.case.study.subcluster.merging.pairing.fluxfrozen.disk,shi:2024.imbh.growth.feedback.survey,rodriguez:balls.of.fire.star.cluster.destruction}, where the measured probability $P_{10}$ for $>10$ such events (a factor $\sim 10^{3-4}$ in total growth) is nearly identical to $P_{2}$ for $>2$, provided (1) massive enough sub-clouds continue to exist and (2) BH growth is not ultimately limited by BH feedback (which depends on the feedback efficacy and sets the BH-$\sigma$ relation; \citealt{hopkins:qso.stellar.fb.together,hopkins:2021.bhs.bulges.from.sigma.sfr,shi:2024.imbh.growth.feedback.survey}) or stellar feedback (which sets the threshold surface density/column for accretion; \citealt{grudic:sfe.cluster.form.surface.density,grudic:max.surface.density,shi:2022.hyper.eddington.no.bhfb}).

This ``leapfrog'' mode requires that the cloud be bound, i.e.\ $\Delta v \lesssim v_{\rm cl}$; otherwise the BH punches through the cloud with negligible capture. This scenario is most relevant for IMBHs and seed BHs in dense, gas-rich environments, as studied in \citet{shi:2024.seed.to.smbh.case.study.subcluster.merging.pairing.fluxfrozen.disk}. It also requires that the host cloud be sufficiently high-$\Sigma$ that its star-formation/destruction timescale is longer than the BH-cloud crossing time, and that the BH-cloud relative velocity is not too super-virial (i.e.\ there must exist clouds/clumps with internal velocities comparable to the BH velocity dispersion).

In this ``slow encounter'' limit, the effective $\Delta v \lesssim v_{\rm cl}$, so we can use the scalings above with $\tilde{v} \sim 1-2$
\begin{align}
\frac{H_{d}}{r_{t}} &\sim {4 \alpha_{\rm cl,\,B}} \sim 1 \ , \\
Q_{d} &\sim \frac{256\pi}{3}\,\alpha_{\rm cl,\,B}^{3} \gg 1 \ , \\
\Delta t_{d} &\sim \frac{\Omega^{-1}}{16 \alpha^{2}_{\rm cl,\,B}} \sim \Omega^{-1}  \ ,  
\end{align}
and
\begin{align}
\dot{m} &\sim  \frac{v_{\rm cl} t_{S}}{4\,R_{\rm cl}} \sim 630 \left( \frac{ \Sigma_{\rm sub} }{10^{4} {\rm M_{\odot} pc^{-2}}} \right)^{3/4} \left( \frac{100\,M_{\odot}}{M_{\rm BH}} \right)^{1/4} \ . 
\end{align}

\subsection{What Happens to Material Which is Disrupted But Not Captured?}

As noted above, in the $M_{\rm BH} \ll M_{\rm cl}$ limit at the nominal ``tidal radius'' $r_{\rm tidal} \sim (M_{\rm cl}/M_{\rm BH})^{1/3} R_{\rm cl}$, where tidal acceleration from the BH become important, we have $v_{K} \sim v_{\rm cl,\,g} (M_{\rm BH}/M_{\rm cl})^{1/3} \ll v_{\rm cl,\,g}$, so cloud material with $\alpha_{\rm cl} \sim 1$ would not be bound to the BH. 

But per \S~\ref{sec:lowmass.bhs:hiacc} because the cloud is turbulent, there is a distribution of velocities: assuming Maxwellian turbulent velocity distributions with maximum coherence-length $\sim R_{\rm cl}$, so the mass with some velocity scales as $dM = M_{\rm cl} (2\pi v_{\rm cl}^{2})^{-3/2} \exp{(-v^{2} / 2 v_{\rm cl}^{2} )} 4\pi v^{2} dv$, then if $\Delta v \sim v_{\rm cl} > v_{K}$ and $v_{K} \ll v_{\rm cl}$, this mass (captured at $r_{\rm tidal}$) can be approximated by $M_{\rm capt}(r \sim r_{\rm tidal}) \sim M_{\rm cl}^{2/3} M_{\rm BH}^{1/3} f(q)$ ($f(q) \equiv 1.6\,q\,\exp{(-q)}$, $q \equiv \Delta v^{2}/2v_{\rm cl}^{2}$). For the global scalings above $q \sim 0.5\,(M_{\rm cl}/M_{0})^{-1/2}$, giving $f(q) \sim 0.5$ for large clouds with $\Delta v \sim v_{\rm cl}$ and a sharp cutoff at $M_{\rm cl} \lesssim 0.03\,M_{0}$. 

For the small clouds in the ISM, with very small BHs, so $M_{\rm BH} \ll M_{\rm cl} \lesssim 0.03\,M_{0}$, at the ``tidal'' radius there is negligible mass bound/captured to the BH, but rather the material at $ r_{\rm capt} \lesssim r \lesssim r_{\rm tidal}$ will be disrupted and unbound into an outflow/stream moving at $\sim \Delta v \gtrsim v_{\rm cl}$. For clouds more massive than $M_{\rm cl} \gtrsim 0.03\,M_{0}$, however (which contain most of the cloud mass and, we argue, dominate the duty cycle), a mass $\sim 0.5\,M_{\rm BH}\,(M_{\rm cl}/M_{\rm BH})^{2/3} \gg M_{\rm BH}$ (but $\ll M_{\rm cl}$) will be captured and circularize at $R \sim r_{\rm tidal} \gg r_{\rm capt}$, with $v_{\rm turb} \sim v_{K}(R)$ (since most of this mass is near the threshold), so it will be (by definition) outside the BH radius of influence, and self-gravitating. This can therefore form an extended, star-forming ``disk,'' dusty ``torus,'' or local ISM on scales outside the ROI, as seen in simulations of similar encounters with these more modest mass ratios \citep{bonnell.rice:2008.gmc.infall.smbh.sf.sims,shi:2024.seed.to.smbh.case.study.subcluster.merging.pairing.fluxfrozen.disk,bardati:2025.superzoom.torus.rad.transfer.modeling,bardati:2026.hot.dog.fif.simulation.obs.comparison.galaxy.starburst}.

\subsection{Long-Term Evolution (Through and Post Encounter)}

First, note that the scalings for this regime implicitly assume that the cloud can \textit{survive} under its own internal evolution physics for of order the ``encounter timescale'' -- i.e.\ the cloud must be able to survive for at least one BH-cloud crossing time before being destroyed by star formation and stellar feedback or other internal processes (or else the accretion would be disrupted before the gas mass we estimated could be accreted). This requires (1) that the cloud-BH crossing time $\sim R_{\rm cl}/\Delta v$ be shorter than the star formation/depletion timescale which tends to be a few cloud internal crossing times $\sim {\rm few} \times R_{\rm cl}/v_{\rm cl}$ (independent of the cloud star formation efficiency $\epsilon_{\rm ff}$; see \citealt{lee:gmc.sfe,grudic:sfe.gmcs.vs.obs,grudic:sfe.cluster.form.surface.density,grudic:max.surface.density,grudic:cluster.properties,guszejnov:gmc.to.protostar.semi.analytic,guszejnov:fire.gmc.props.vs.z,benincasa:2020.gmc.lifetimes.fire}). But that is relatively easy to satisfy since $v_{\rm cl} \lesssim {\rm few}\,\Delta v$. More importantly, the rapidly-growth scenario (2) requires that during that time, an $\mathcal{O}(1)$ fraction of the gas remains bound, and not be efficiently unbound or accelerated to much higher velocities, especially in dense regions/clumps/sub-clouds where stars are actually forming. Otherwise, as shown in e.g.\ \citep{shi:2022.hyper.eddington.no.bhfb,shi:2024.seed.to.smbh.case.study.subcluster.merging.pairing.fluxfrozen.disk,shi:2024.imbh.growth.feedback.survey} and other studies \citep{lupi:2024.highz.smbh.growth.supereddington,partmann:2024.nuclear.star.clusters.critical.shelter.focus.accretion.drive.centering.bhs.in.dwarfs,rantala:2026.cluster.bifrost.ketju.imbh.sims.form.efficiently.in.dense.low.metallicity.clusters.as.expected}, the ``leap-frog'' scenario (\S~{sec:lowmass.bhs:hiacc}) is completely suppressed (the probability of encountering a dense clump with low relative velocity is suppressed by orders of magnitude), and only the low-$\dot{M}$ ``turbulent Bondi-Hoyle''-like channel is further suppressed by the larger $\delta v_{\rm turb}$ driven by feedback. This requires high-surface-density clouds, $\Sigma_{\rm cl} \gg 1000\,{\rm M_{\odot}\,pc^{-2}}$, at which point gravity (force $\sim G M_{\rm cloud} M_{\rm gas}/R^{2} \sim G \Sigma_{\rm cl} M_{\rm gas}$) overwhelms feedback (force $\dot{P} \sim L/c \sim 1200\,L_{\odot}/(M_{\odot}\,c) M_{\ast}$) unless $M_{\ast} \gtrsim M_{\rm gas}$, as shown in many observational and theoretical studies of cluster-forming clouds \citep{fall:2010.sf.eff.vs.surfacedensity,murray:2010.sfe.mw.gmc,2017MNRAS.471.4844G,2018ApJ...859...68K,grudic:sfe.cluster.form.surface.density,grudic:max.surface.density,hopkins:2021.bhs.bulges.from.sigma.sfr,shi:2022.hyper.eddington.no.bhfb,shi:2024.imbh.growth.feedback.survey}. This is quite reasonable compared to typical dense gas complexes in starburst nuclei or gas-rich high-redshift galaxies, which extend up to $\Sigma_{\rm cl} \sim 10^{5}\,{\rm M_{\odot}\,pc^{-2}}$, but much larger than the typical $\sim 70\,{\rm M_{\odot}\,pc^{-2}}$ of Solar-neighborhood ISM clouds. 

Once the cloud-BH interaction ceases (after a crossing time $\sim R_{\rm cl}/\Delta v$), the disk is no longer replenished. 
If the mass supply were constant, then instantaneously removed, Eq.~\ref{eqn:mdot.general} implies the disk would deplete and $\dot{M}_{d}$ would decay roughly exponentially on the timescale $\Delta t_{d}$. 
However, because the BH is capturing from an effectively infinite, turbulent medium, the distribution of specific energies of material captured (${\rm d} M_{\rm capt} / {\rm d} E^{\rm spec}_{\rm capt}$) must be continuous through $E^{\rm spec}_{\rm capt}=0$. As a result, the long-time accretion rate once the supply is ``cut off'' (e.g.\ the BH exits dense regions of the cloud entirely) should approximately follow the usual $\dot{M} \propto t^{-5/3}$ relation \citep{phinney:1989.tidal.disruptions.galactic.center.bh.53.lightcurve}, with the characteristic timescale set by $\Delta t_{d} \sim \Omega^{-1}$ at $r_{t}$.
Since this is a power-law, it will dominate the long-term decay over the exponential case. 
The even duration up to this point is determined by the time to cross the cloud, $\Delta t_{\rm event} \sim R_{\rm cl}/\Delta v \sim (M_{\rm cl}/M_{0})^{1/2}\,\Omega_{0}^{-1}$.

\section{Comparable-Mass Encounters: $M_{\rm BH} \sim M_{\rm cl}$}
\label{sec:comparable.mass}

Now consider the regime where $M_{\rm cl} \sim M_{\rm BH}$ or $M_{\rm cl} \gtrsim M_{\rm BH}$ (up to factors of a few). This is the regime relevant for massive BHs ($M_{\rm BH} \sim 10^{6-9}\,M_{\odot}$) encountering the most massive cloud complexes in gas-rich galactic nuclei, as well as the ``effective'' sub-cloud encounter scenario from \S~\ref{sec:lowmass.bhs:hiacc} (where a rare slow encounter has $M_{\rm sub} \sim M_{\rm BH}$, i.e.\ $\tilde{v} \sim 1$).

In this regime, using the scalings above with $\tilde{v} \sim 1-2$ (cloud bound to BH potential), the tidal/capture radius is $r_{t} \sim R_{\rm cl}\,(M_{\rm BH}/M_{\rm cl})\,\tilde{v}^{-2} \sim R_{\rm cl}$, so the capture radius is comparable to the cloud size. The captured mass $M_{\rm capt} \sim M_{\rm cl}$ (most of the cloud mass is captured), and the Keplerian velocity at the capture radius is $v_{K}(r_{t}) \sim v_{\rm cl}$. The capture cross-section is $\sigma_{\rm BH-cl} \sim \pi R_{\rm cl}^{2}$, comparable to the geometric cloud area.

These intermediate expressions follow from the general scalings in \S~\ref{sec:setup} and \S~\ref{sec:lowmass.bhs}. The key scalings for the resulting disk are:
\begin{align}
\frac{H_{d}}{r_{t}} &\sim {4 \alpha_{\rm cl,\,B}} \sim \mathcal{O}(1) \ , \\
Q_{d} &\sim \frac{256\pi}{3}\,\alpha_{\rm cl,\,B}^{3} \left(\frac{M_{\rm cl}}{M_{\rm BH}}\right)^{2} \gtrsim 1 \ , \\
\Delta t_{d} &\sim \frac{\Omega^{-1}}{16 \alpha^{2}_{\rm cl,\,B}} \sim \Omega^{-1}  \ ,
\end{align}
and the Eddington-normalized accretion rate:
\begin{align}
\dot{m} &\sim  \frac{v_{\rm cl} t_{S}}{4\,R_{\rm cl}} \sim 7\,\alpha_{\rm cl,\,B} \left( \frac{ \Sigma_{\rm cl} }{100\,{\rm M_{\odot}\,pc^{-2}}} \right)^{3/4} \left( \frac{M_{\rm cl}}{10^{7}\,{\rm M_{\odot}}} \right)^{-1/4} \left( \frac{2}{\tilde{v}} \right) \ .
\end{align}

So in the regime where the cloud is bound to the BH ($\Delta v \lesssim v_{\rm cl}$), high-$\dot{m}$ accretion is essentially guaranteed. For high surface-density complexes ($\Sigma_{\rm cl} \sim 10^{4-5}\,{\rm M_{\odot}\,pc^{-2}}$, as can be realized in dense nuclear star-forming regions and high-redshift starburst galaxies), the accretion can be highly super-Eddington, with $\dot{m}$ reaching several hundred to a thousand.

In this regime, the gravity of the BH is not totally dominant over the self-gravity of the captured material at the outermost radii ($Q_{d}$ can be only modestly above unity), which suggests likely $\mathcal{O}(1)$ eccentricity of the circularized material (which would also inevitably arise because of the arbitrary eccentricity of the ``captured'' orbit), significant external gravitational perturbations, and sloshing/warping modes. These perturbations can themselves drive vigorous angular momentum transport \citep{shlosman:bars.within.bars,gammie:2001.cooling.in.keplerian.disks,hopkins:zoom.sims,hopkins:inflow.analytics,hopkins:superzoom.disk,kaaz:2022.grmhd.sims.misaligned.acc.disks.spin}, further sustaining the magnetically-dominated, turbulent accretion state.

\subsection{Magnetic Coherence and Field Sustainability}

In this regime, because $r_{t} \sim R_{\rm cl}$, the coherence length of the captured magnetic field $\lambda_{\rm coh} \sim R_{\rm cl} \sim r_{t}$. This means the toroidal field is coherent on scales comparable to the disk radius, which is the most favorable configuration for sustaining the field against both Parker instability losses and turbulent dissipation. Comparing the advective (accretion) timescale to the Parker instability timescale (as in \S~\ref{sec:lowmass.bhs:lowacc}): $\Delta t_{d}/\Delta t_{\rm Parker} \sim 1/(500 \times 16\,\alpha_{\rm cl,\,B}^{2}) \ll 1$ for any reasonable $\alpha_{\rm cl,\,B}$. Similarly, the turbulent flux transport time is also long compared to the advection time. This suggests that the initial toroidal-field-dominated state could be self-sustaining throughout the accretion event.

\section{Very High-Mass BHs: $M_{\rm BH} \gg M_{\rm cl}$}
\label{sec:highmassBH}

Some behaviors should change qualitatively when the cloud mass is much smaller than the BH mass ($M_{\rm cl} \ll M_{\rm BH}$). The capture/Bondi radius is still $r_{\rm capt} = GM_{\rm BH}/v_{e}^{2} = R_{\rm cl}\,(M_{\rm BH}/M_{\rm cl})\,\tilde{v}^{-2}$. Comparing this to $R_{\rm cl}$ or the tidal radius $r_{\rm tidal} = (M_{\rm BH}/M_{\rm cl})^{1/3} R_{\rm cl}$ shows that $r_{\rm capt} \ll R_{\rm cl}$ or $r_{\rm capt} \ll r_{\rm tidal}$ requires $\tilde{v} \gg (M_{\rm BH}/M_{\rm cl})^{1/2}$ or $\tilde{v} \gg (M_{\rm BH}/M_{\rm cl})^{1/3}$, respectively. But using the scalings above ($\tilde{v} \rightarrow (M_{\rm cl}/M_{0})^{-1/4}$ for small $M_{\rm cl}$), if $M_{\rm cl} \ll M_{\rm BH}$ (the case of interest here), this requires $M_{0} \gg M_{\rm BH}\,(M_{\rm BH}/M_{\rm cl})$ or $M_{0} \gg M_{\rm BH}\,(M_{\rm BH}/M_{\rm cl})^{1/3}$. So $r_{\rm capt} \ll r_{\rm tidal}$ or $R_{\rm cl} \ll r_{\rm tidal}$ is, in practice, very difficult to satisfy for any of the conditions of interest (it requires $M_{\rm cl} \ll M_{\rm BH}$, but also $M_{0} \gg M_{\rm BH}$, while sampling clouds at the extreme low end of the cloud mass function), and in those conditions would produce extremely little accretion owing to (1) the suppression of the accretion by $\tilde{v}$ (in standard Bondi-like fashion), and (2) the small cloud mass compared to both the BH and other clouds. So although technically possible, this is a much less interesting and less-relevant regime (and $R_{\rm cl} \ll r_{\rm tidal}$ is even more difficult to achieve). 

In other words, for small clouds, the salient circularization/disk radius is the tidal disruption radius, which is generally now \textit{interior} to the capture radius (opposite the large-cloud case). Therefore we will parameterize $r_{t}$ as: $r_{t} \approx \xi (M_{\rm BH}/M_{\rm cl})^{1/3} R_{\rm cl}$ with 
\begin{align}
\xi \equiv \left[ 1 + \left( \frac{M_{\rm cl}}{M_{\rm BH}}\right)^{2/3}\,\tilde{v}^{2} \right]^{-1} \ , 
\end{align}
 which interpolates between $r_{t} \sim r_{\rm tidal}$ ($\xi\sim1$) and $r_{t} \sim r_{\rm capt}$ ($\xi \ll 1$, $\tilde{v} \gg (M_{\rm BH}/M_{\rm cl})^{1/3}$), but we will generally have $\xi \sim 1$.
We then obtain: 
\begin{align}
r_{t} &\sim \xi \, \left( \frac{M_{\rm BH}}{M_{\rm cl}} \right)^{1/3} R_{\rm cl} \gg R_{\rm cl} \ , \\
v_{K} &\sim \xi^{-1/2} \left( \frac{M_{\rm BH}}{M_{\rm cl}} \right)^{1/3} v_{\rm cl} \gg v_{\rm cl} \ , \\
\sigma_{\rm BH-cl} &\sim \pi r_{t}^{2} \gg \pi R_{\rm cl}^{2} \ , \\ 
M_{\rm capt} &\sim M_{\rm cl} \ . 
\end{align}

Since the cloud can disrupt at $r_{t} \gg R_{\rm cl}$, most of the mass will initially form a toroidal geometry with vertical width $\ell_{z,\,d} \sim H_{d} \sim (\delta v_{z} /v_{K})\,r_{t}$ (for some vertical support $\delta v_{z}$) and radial ``width'' $\ell_{R,\,d} \sim \Delta R  \sim (\delta v_{R}/v_{K})\,r_{t}$ ($\delta v_{R} \equiv \zeta_{R} \delta v_{z}$), sheared into azimuthal extent $\ell_{\phi,\,d} \sim 2\pi r_{t}$. This implies the ratio of disk-to-cloud density $\rho_{d}/\langle \rho_{\rm cl} \rangle  =  (M_{\rm cl} / 4\pi R_{\rm cl}^{3}/3) / (M_{\rm cl} / 2 \pi r_{t} \Delta R 2 H_{d}) \sim R_{\rm cl}^{3}/(3 \zeta_{R} H_{d}^{2} r_{t})$. 
For thermal support, cooling implies $c_{s,\,d} \sim c_{s,\,\rm cl} \ll v_{\rm cl}$ is small compared to either turbulence or magnetic support (calculated in the following). 
If magnetic support is dominant, then $\zeta_{R} \sim 1$ and $\delta{v}_{z} \sim \delta v_{R} \sim v_{A,\,\phi} \sim v_{A,\,\rm cl} (B_{\phi,\,d}/B_{\phi,\,0}) / \sqrt{\rho_{d} /\langle \rho_{\rm cl} \rangle} \sim v_{\rm cl} (3 \alpha_{B,\,\rm cl} \xi)^{1/2} (M_{\rm BH}/M_{\rm cl})^{1/6}\,(R_{\rm cl}/H_{d}^{A})$. 
This implies $H_{d}^{A}/r_{t} \sim (3 \alpha_{B,\,\rm cl})^{1/4} \,(M_{\rm BH}/M_{\rm cl})^{-1/4}$. 
If, on the other hand, turbulence dominates, the maximum support comes if we assume ``adiabatic'' enhancement (effective index $P_{\rm turb} \propto \rho^{5/3}$) of $v_{\rm turb}$ and isotropic motions, so $\delta v_{z,\,d} \sim \delta v_{R,\,d} \sim v_{\rm turb,\,d} \sim v_{\rm turb,\,\rm cl} (\rho_{d}/\langle \rho_{\rm cl} \rangle)^{1/3} \sim v_{\rm cl}\,(3 \xi\,[H_{d}/R_{\rm cl}]^{2} [M_{\rm BH}/M_{\rm cl}]^{1/3})^{-1/3}$. Consistency with $v_{\rm turb} \sim (H_{d}/r_{t}) v_{K}$ then implies $v_{\rm turb} \sim 3^{-1/5} \xi^{-3/10} (M_{\rm BH}/M_{\rm cl})^{-2/5}\,v_{\rm cl}$, $H_{d}^{\rm turb}/r_{t} \sim 3^{-1/5} \xi^{-3/10} (M_{\rm BH}/M_{\rm cl})^{-2/5}$. 
The ratio of $H_{d}^{\rm A}/H_{d}^{\rm turb} \sim 3^{9/20} \alpha_{B,\,\rm cl}^{1/4} (M_{\rm BH}/M_{\rm cl})^{3/20} \xi^{3/10} \sim 1$ is extremely weakly-dependent on BH mass and encounter properties. Since the turbulent case was explicitly an upper-limit, we therefore more conservatively assume the $H_{d}^{\rm A}$ case, giving: 
\begin{align}
\frac{H_{d}}{r_{t}} &\sim ({3\alpha_{\rm cl,\,B}})^{1/4}\, \left(\frac{M_{\rm BH}}{M_{\rm cl}} \right)^{-1/4} \ , \\
B_{\phi,\,d} &\sim \frac{B_{\rm cl}}{\sqrt{3 \alpha_{\rm cl,\,B}} \xi^{2}} \,\left(\frac{M_{\rm BH}}{M_{\rm cl}} \right)^{-1/6}  \sim  B_{\phi,\,\rm cl} \ , \\
B_{z,\,d} &\sim B_{R,\,d} \sim \frac{B_{\rm cl}}{2\pi ({3 \alpha_{\rm cl,\,B}})^{1/4} \xi^{2}} \,\left(\frac{M_{\rm BH}}{M_{\rm cl}} \right)^{-5/12} < B_{\phi,\,d} \\
Q_{d} &\sim 2 \sqrt{3\alpha_{\rm cl,\,B}} \left(\frac{M_{\rm BH}}{M_{\rm cl}} \right)^{1/2} \gg 1 \ , \\
\Delta t_{d} &\sim \Delta t_{\rm event} \sim \frac{\Omega^{-1}}{\sqrt{3\alpha_{\rm cl,\,B}}} \left(\frac{M_{\rm BH}}{M_{\rm cl}} \right)^{1/2} \ , \\
\dot{M}_{d} &\sim \frac{\sqrt{3\alpha_{\rm cl,\,B}}}{\xi^{3/2}} M_{\rm cl} \frac{v_{\rm cl}}{R_{\rm cl}} \,\left(\frac{M_{\rm BH}}{M_{\rm cl}} \right)^{-1/2}\ .
\end{align}

We also obtain a crude estimate of time-averaged rates, under the assumption that we are moving through a random field (i.e.\ ignoring capture of clouds at $r_{\rm capt} \gg r_{\rm tidal}$ which then migrate more slowly via resonant interactions or dynamical friction in a sort of gas-cloud ``loss cone refilling'' process more analogous to TDEs):
\begin{align}
\frac{d\dot{N}_{\rm events}}{d\ln{M_{\rm cl}}} &\sim \frac{(2-\gamma_{\rm cl}) \xi^{2}}{\psi} \Omega_{0} \left( \frac{M_{\rm BH}}{M_{0}} \right)^{2/3} \left( \frac{M_{\rm cl}}{M_{0}} \right)^{4/3-\gamma_{\rm cl}}  \ , \\
\frac{d{f}_{\rm events}}{d\ln{M_{\rm cl}}} &\sim \frac{(2-\gamma_{\rm cl}) Q_{0}^{1/2} \xi^{7/2}}{(3\alpha_{\rm B,\,cl})^{1/2} \psi^{3/2}} \left( \frac{M_{\rm BH}}{M_{0}} \right)^{7/6} \left( \frac{M_{\rm cl}}{M_{0}} \right)^{13/12-\gamma_{\rm cl}}  \ , \\
\frac{d\dot{M}_{\rm events}}{d\ln{M_{\rm cl}}} &\sim \frac{(2-\gamma_{\rm cl}) \xi^{2}}{\psi} M_{0} \Omega_{0} \left( \frac{M_{\rm BH}}{M_{0}} \right)^{2/3} \left( \frac{M_{\rm cl}}{M_{0}} \right)^{7/3-\gamma_{\rm cl}}  \ .
\end{align}

Taking some plausible $\mathcal{O}(1)$ numbers and $\gamma_{\rm cl} \sim 7/4$, this gives for the relevant integrals:
\begin{align}
\dot{N}_{\rm events} &\sim \frac{3}{5 \psi} \frac{M_{\rm BH}^{2/3}}{M_{\rm cl,\,min}^{5/12} M_{0}^{1/4} } \Omega_{0} \ , \\
f_{\rm events} &\sim \frac{M_{\rm BH}}{6 \psi^{3/2} \sqrt{M_{\rm cl,\,min} M_{0}}} \ , \\
\dot{M}_{\rm events} &\sim \frac{3}{7 \psi} \Omega_{0} \frac{M_{\rm BH}^{2/3} {\rm MIN}[M_{0},\,M_{\rm BH}]^{7/12}}{M_{0}^{1/4}}  \ .
\end{align}
Setting $f_{\rm events} \sim 1$ gives $M_{\rm min}/M_{0} \sim (M_{\rm BH}/6\,M_{0})^{2}$, so if $M_{\rm BH} \lesssim M_{0}$, the duty cycle is of order unity. If $f_{\rm events} > 1$, events overlap, which is physically reasonable and simply means that multiple cloud disruption events are ongoing simultaneously.

\begin{figure*}
\centering
\includegraphics[width=1\textwidth]{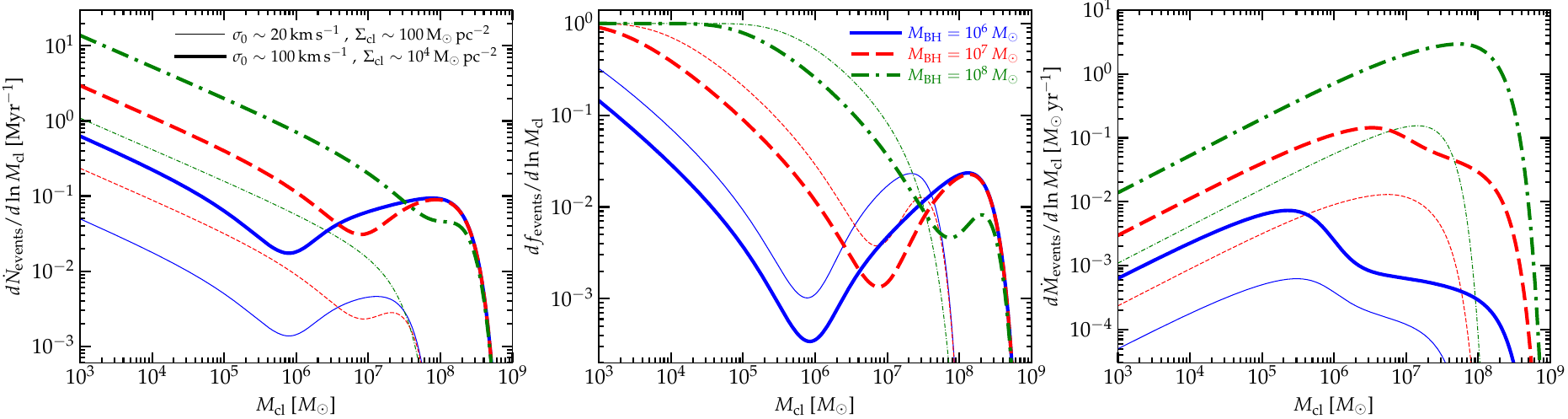}
\caption{Cloud mass spectrum-weighted rates as a function of cloud mass $M_{\rm cl}$, for three BH masses $M_{\rm BH} = 10^{6}$ (blue), $10^{7}$ (red), and $10^{8}\,M_{\odot}$ (green), assuming a cloud mass function with $\gamma_{\rm cl}=7/4$, $M_{0}=10^{8}\,M_{\odot}$, $\Sigma_{0} = 100\,M_{\odot}\,{\rm pc^{-2}}$. {\em Left:} Differential encounter rate $d\dot{N}/d\ln{M_{\rm cl}}$. {\em Center:} Differential time-averaged mass accretion rate $d\dot{M}/d\ln{M_{\rm cl}}$. {\em Right:} Differential duty cycle $df/d\ln{M_{\rm cl}}$. In all cases, the transition between the $M_{\rm cl} > M_{\rm BH}$ and $M_{\rm cl} < M_{\rm BH}$ regimes is visible. The mass accretion rate is dominated by clouds of mass $M_{\rm cl} \sim M_{\rm BH}$, while the event rate is more broadly distributed.}
\label{fig:rates.spectrum}
\end{figure*}

\subsection{Magnetic Field Coherence and Sustainability}

A qualitative difference from the $M_{\rm cl} \gtrsim M_{\rm BH}$ case is that here the magnetic field coherence length is limited to $\lambda_{\rm coh} \sim R_{\rm cl} \ll r_{t}$. This means that the toroidal field, while still dominant over $B_{z}$ and $B_{R}$ in the initial disk, may be more vulnerable to turbulent dissipation (since turbulence on scales $\sim R_{\rm cl}$ can mix and reconnect the field on those scales).

Using the same viscous scaling for $\dot{M}$, we obtain $\dot{M}_{d} \sim 3\,\alpha_{\rm cl,\,B}\,v_{\rm cl}^{3}/G$, which gives $\dot{M}_{d} \propto M_{\rm cl}^{3/4}$ for the population of clouds. The Eddington-normalized rate is:
\begin{align}
\dot{m} \sim &0.02 \left( \frac{\alpha_{\rm cl,\,B}}{0.1}\right) \left(\frac{\Sigma_{\rm cl}}{100\,{\rm M_{\odot}\,pc^{-2}}}\right)^{3/4} \\ 
\nonumber & \times \left(\frac{M_{\rm cl}}{0.01\,M_{\rm BH}}\right)^{3/4} \left(\frac{M_{\rm BH}}{10^{8}\,M_{\odot}}\right)^{-1/4}\ .
\end{align}
This is the outer $\dot{m}$; it can drop more rapidly inward if there are outflows or other barriers to accretion.

The depletion-to-dynamical time ratio is $\Delta t_{d}/\Omega^{-1} \sim \alpha_{\rm cl,\,B}^{-1}(M_{\rm BH}/M_{\rm cl})^{2/3}$. If this exceeds $\sim 500$ (the Parker instability flux-loss timescale in units of $\Omega^{-1}$), the toroidal field can be lost before being replenished. For $\alpha_{\rm cl,\,B} \sim 0.1$, this threshold corresponds to $M_{\rm cl} \lesssim 0.004\,M_{\rm BH}$, i.e.\ $\dot{m} \lesssim 0.01\,(\Sigma_{\rm cl}/100\,{\rm M_{\odot}\,pc^{-2}})^{3/4}\,(M_{\rm BH}/10^{8}\,M_{\odot})^{-1/4}$.

For realistic astrophysical parameters, this low-$\dot{m}$, less-magnetized regime is only accessed when the galaxy's gas fraction is low and $M_{0} \ll M_{\rm BH}$, which requires $f_{\rm gas} \ll 0.1$ (for a BH already on the $M$-$\sigma$ relation). Given realistic limits on the stellar surface density, $\Sigma_{\rm cl}$ cannot exceed $\sim 1000\,{\rm M_{\odot}\,pc^{-2}}$ in this limit, implying even $\dot{m} \ll 0.05$ at outer radii must be very small. This is broadly consistent with the transition to radiatively-inefficient accretion flows observed in low-luminosity AGN, and corresponds to $H/R \sim 0.05$ (for $\alpha_{\rm cl,\,B} \sim 0.1$). So it again corresponds to the regime where our assumptions about Parker modes would not apply in the first place, because a spherical/convective flow would likely emerge owing to inefficient cooling in the flow.

\section{High-$\beta$ Initial ``Clouds'': Stellar Binaries \&\ Tidal Disruption}
\label{sec:highbeta}

\subsection{Reminder of Low-$\beta$ Predictions}

For ISM clouds with $\beta_{\rm cl} \ll 1$ (i.e.\ $\alpha_{\rm thermal} \ll \alpha_{\rm turb} \sim \alpha_{B} \sim 1$), the scalings derived above robustly predict $\beta_{d} \lesssim \beta_{\rm cl} \ll 1$ in both the low and high-BH-mass limits. In the high-$M_{\rm cl}$ regime (small BHs), $\beta_{d}/\beta_{\rm cl} \sim \tilde{v}^{4}/(16\alpha^{2}_{\rm cl,\,B})$, which is not very large unless $\tilde{v}$ is huge (very difficult to achieve) or $\alpha_{\rm cl,\,B} \ll 0.1$ (which would violate the initial assumptions). In the low-$M_{\rm cl}$ regime (large BHs, \S~\ref{sec:highmassBH}), $\beta_{d}/\beta_{\rm cl} \sim 1/(3\,\xi\,\sqrt{M_{\rm BH}/M_{\rm cl}}) \ll 1$. 

Meanwhile, in both small-and-large BH limits, the initial \Alf\ Mach number of the disk turbulence is predicted to be modest, $\mathcal{M}_{A,\,d} \equiv v_{\rm turb,\,d}/v_{A,\,d} \sim 0.5-2$. This plus $\beta_{d} \ll 1$ means the sonic Mach number should be large, $\mathcal{M}_{s,\,d} \gg 1$. In other words, the low-$\beta$, trans-\Alf{ic}, supersonic, magnetically-dominated state is a robust outcome of ISM cloud accretion.

\subsection{The High-$\beta$ Limit: Stellar-Origin Material}

The situation is qualitatively different if $\beta_{\rm cl} \gg 1$ in the initial ``cloud.'' This is the relevant limit for accreting stellar material, as in tidal disruption events (TDEs), X-ray binaries (XRBs), cataclysmic variables (CVs), and other accreting stellar-mass compact objects. The incoming material has $\beta_{\rm cl} \gg 1$, with $\alpha_{\rm thermal} \sim 1 \gg (\alpha_{\rm turb},\,\alpha_{B})$, and is optically thick to its own cooling radiation so the sound speed behaves adiabatically ($c_{s} \propto \rho^{1/3}$).

Consider the two limits: (a) $M_{\rm BH} \ll M_{\rm cl}$, and (b) $M_{\rm BH} \gg M_{\rm cl}$. In limit~(a), taking our previously-derived scalings plus flux-freezing immediately shows that the problem reduces to the true classical Bondi-Hoyle problem (or pure Bondi if $\tilde{v}$ is small). One obtains the standard prediction: post-accretion/shock, $c_{s,\,d} \sim v_{e}$, $\rho_{d} \sim \rho_{0}$, $H_{d} \sim r_{t}$, $r_{t} \sim r_{\rm Bondi} \sim G M_{\rm BH}/v_{e}^{3}$, $B_{\phi,\,d} \sim 2\pi\,B_{z,\,d} \sim 2\pi\,B_{R,\,d} \sim B_{\rm cl}$, so $\beta_{d} \gtrsim \beta_{\rm cl} \gg 1$. In other words, one obtains the expected result of spherical Bondi accretion with weak ``initial'' magnetic support, modest (sub or trans-sonic) turbulence, and a thermally-supported, approximately spherical flow.

In limit~(b), which is the relevant case for TDEs, solving the same scalings gives $H_{d}/r_{t} \sim (M_{\rm cl}/M_{\rm BH})^{2/5} 3^{-1/5} \xi^{-3/10}$, $B_{\phi,\,d} \sim 3^{2/5}\,(M_{\rm cl}/M_{\rm BH})^{-2/15}\xi^{-7/5} B_{\rm cl}$,
$B_{z,\,d} \sim B_{R,\,d} \sim (3^{1/5}/2\pi)\,(M_{\rm cl}/M_{\rm BH})^{4/15}\xi^{-17/10} B_{\rm cl} < B_{\phi,\,d}$, and
$\beta_{d}/\beta_{\rm cl} \sim 3^{-9/5} \xi^{-6/5} (M_{\rm cl}/M_{\rm BH})^{3/5}$.
If we ignore shock and accretion heating, we predict $\beta_{d} < \beta_{\rm cl}$; but to reach the magnetically-dominated limit $\beta_{d} \lesssim 1$ would require $\beta_{\rm cl} \lesssim 3^{9/5} \xi^{6/5} (M_{\rm cl}/M_{\rm BH})^{-3/5}$. In extreme TDE scenarios (where $M_{\rm cl} \ll M_{\rm BH}$ by many orders of magnitude), this could in principle be satisfied even for stellar initial $\beta_{\rm cl}$. However, this ignores the dominant effect: if the material is shock-heated to $c_{s} \sim v_{K}$ at the disruption radius and cannot rapidly cool (as expected for adiabatic material), then $H_{d}/r_{t} \sim 1$, $c_{s,\,d} \sim v_{K}$, $\rho_{d} \sim \xi^{-3} (M_{\rm cl}/M_{\rm BH})\,\rho_{\rm cl}$, and $\beta_{d} \sim (M_{\rm BH}/M_{\rm cl})\,\beta_{\rm cl} \gg \beta_{\rm cl}$. So TDEs and other stellar-origin accretion events will {not} be naively expected to produce magnetically-dominated initial disk states. This is consistent with the well-known result that TDE disks are expected to be thermally-supported and radiation-pressure dominated, with magnetic fields playing a subdominant role in the initial configuration (though they may be amplified subsequently by the MRI and dynamo action).

\subsection{An Intermediate Case: Hot Halos with Thermal Instability}

A more complicated case arises when $\beta_{\rm cl} \gg 1$ initially, but the cooling time becomes short compared to the capture/circularization time at some intermediate radius. This is physically relevant for, e.g., Bondi accretion from a hot ($\gtrsim 10^{7}\,$K), X-ray emitting atmosphere on $\gtrsim$\,kpc scales in systems like M87. As the accretion flow becomes denser at smaller radii, it becomes thermally unstable and cools strongly, producing molecular gas at $\sim 10$\,K \citep{li.bryan:2014.cooling.flow.cold.gas,li:2015.cooling.agn,tremblay:2016.m87.cold.gas,prasad:2017.stochastic,tremblay:2018.fountain,olivares:2019.cold.gas.clusters,ressler:2020.sgra.magnetic}. The thermal pressure drops by $\sim 6$ orders of magnitude while the magnetic pressure (being flux-frozen) remains roughly constant or increases, producing a low-$\beta$ cold gas phase.

In this case, there is no trivial scale-free expectation, but one can apply the intuition from the various regimes above to different radial ranges. Multi-scale simulations \citep{cho:2023.multiscale.accretion.sims.bondi.inflow.model,cho:2024.multizone.grmhd.sims.bondi.flow.lowaccrate,guo:2024.fluxfrozen.disks.lowmdot.ellipticals,guo:2025.toroidal.disk,wang:2025.hypermagnetized.circumbinary.disk.flux.frozen.cavity.to.pc.scales} show precisely this behavior: a transition from a Bondi-like flow with $\beta \gg 1$ at large radii, to a magnetically-dominated $\beta \ll 1$ cold disk produced by tidal capture of cool clouds from thermal instabilities at intermediate radii, and then a transition back to a spherical, virialized flow at small radii where $\dot{m} \ll 1$ implies very inefficient cooling in the inner accretion disk. 

\section{Cosmological Context}
\label{sec:cosmo}

\begin{figure*}
\centering
\includegraphics[width=0.9\textwidth]{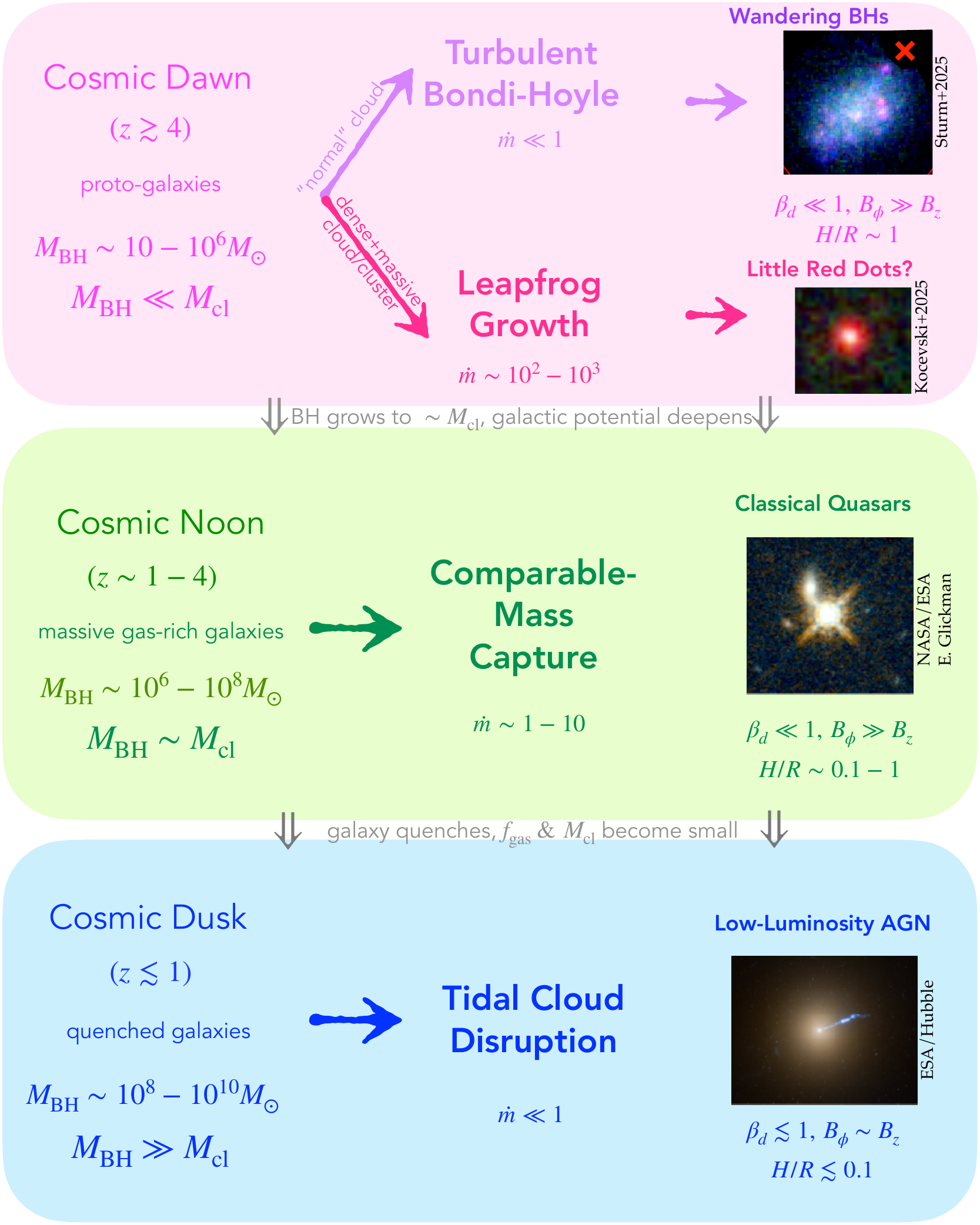}
\caption{Schematic overview of cloud accretion regimes across cosmic time. Each row corresponds to one of the three mass-ratio regimes derived in this paper, mapped to the galaxy populations and cosmic epochs where it is most relevant for a massive galaxy today. {\em Top:} Low-mass BH seeds in dwarf and proto-galaxies at high-$z$. Most seeds wander with negligible growth (turbulent Bondi-Hoyle), but those embedded in sufficiently dense cloud complexes can undergo ``leapfrog'' hyper-Eddington growth, potentially producing ``little red/blue dot''-like objects. {\em Middle:} Comparable-mass cloud capture in gas-rich, $L^{\ast}$ galaxies at intermediate redshifts, produces near-Eddington accretion rates with the maximum in absolute $\dot{M}$, corresponding to the classical quasar population. {\em Bottom:} Tidal disruption of small clouds (from the ISM or thermal instability in cooling halos) in massive, quenched ellipticals at low redshifts, producing LLAGN or RIAF-like accretion. In all regimes where cold clouds are accreted, the resulting disk is toroidally-magnetically-dominated ($\beta_{d} \ll 1$, $B_{\phi} \gtrsim B_{z}$), but how extreme this is can vary.}
\label{fig:cartoon}
\end{figure*}

The scalings derived above allow us to sketch a picture of how cloud accretion operates across cosmic time and in different environments (Fig.~\ref{fig:cartoon}). The three regimes of cloud-to-BH mass ratio map naturally onto three broad classes of galaxies and cosmic epochs. We consider each in turn.

\subsection{Seed BHs in Dwarfs and Early Galaxies}
\label{sec:cosmo:seeds}

For the lowest-mass BH seeds (stellar-mass to $\sim 10^{3}\,M_{\odot}$), whether formed from Population~III remnants, direct collapse, or dynamical processes in dense star clusters \citep{greene:2020.imbh.review,inayoshi:2020.bh.seed.review,lupi:2024.highz.smbh.growth.supereddington}, the BH is always in the $M_{\rm BH} \ll M_{\rm cl}$ regime (\S~\ref{sec:lowmass.bhs}) with respect to all but the smallest ISM structures. This regime corresponds to dwarf galaxies at low redshift and proto-galaxies at early times. For ``typical'' encounters, the BH is in the turbulent Bondi-Hoyle regime (\S~\ref{sec:lowmass.bhs:lowacc}), with $\tilde{v} \gg 1$ owing to large random velocities relative to the ambient ISM. In this regime, $\dot{m} \ll 1$, and the BH essentially does not grow. This is consistent with the well-known result from cosmological simulations with resolved ISM physics and stellar feedback that low-mass BH seeds ``wander'' in their host galaxies and fail to accrete efficiently \citep{habouzit:2017.bh.growth.cosmosims.suppressed.by.sne.fb,2018ApJ...857L..22T,pfister:2019.bh.dynamics.dwarfs,ma:2021.fire.smbh.growth.dwarfs,wellons:2022.smbh.growth,weller:2022.imbh.in.dwarf.galaxies.wandering}. Observationally, this is consistent with recent detections of off-nuclear or ``wandering'' massive BHs in dwarf galaxies \citep[e.g.][]{reines:2020.wandering.bhs,mezcua:2024.imbh.dwarfs}, which suggest that many low-mass BHs remain dynamically decoupled from their host galaxy centers.

\subsubsection{Leapfrog Growth and Little Red Dots}
\label{sec:cosmo:leapfrog}

However, if and when sufficiently dense and massive clouds form --- for example, at intermediate metallicities ($Z \sim 10^{-2}\,Z_{\odot}$) where cooling is efficient but stellar feedback is not yet maximally disruptive, as in proto-massive globular cluster environments --- the conditions for the ``leapfrog'' growth mode (\S~\ref{sec:lowmass.bhs:hiacc}) can be met. In this case, hierarchical sub-cloud capture events drive runaway BH growth at rates up to hyper-Eddington ($\dot{m} \gg 1$), allowing the BH to leap from $\lesssim 10^{2}\,M_{\odot}$ to intermediate or even SMBH masses ($\gtrsim 10^{5}\,M_{\odot}$) within $\lesssim$\,Myr, as seen in explicit simulations of this process \citep{shi:2022.hyper.eddington.no.bhfb,shi:2024.seed.to.smbh.case.study.subcluster.merging.pairing.fluxfrozen.disk,shi:2024.imbh.growth.feedback.survey}. Similar rapid BH growth in dense stellar/gaseous environments has been studied in N-body and hydrodynamic simulations by \citet{rizzuto:2021.imbh.cluster,rizzuto:2023.bh.growth.dense.clusters,escala:2021.bh.growth.nuclear.clusters,bruel:2023.fire.starforge.cluster.models.to.ligo.bbh.merger.predictions,lupi:2024.highz.smbh.growth.supereddington,lupi:2024.highz.bhs.agn.are.super.eddington.quasars,partmann:2024.nuclear.star.clusters.critical.shelter.focus.accretion.drive.centering.bhs.in.dwarfs,mukherjee:2024.nuclear.star.clusters.needed.smbh.sinking.and.tidal.allows.mergers.binary.pairing,rantala:2026.cluster.bifrost.ketju.imbh.sims.form.efficiently.in.dense.low.metallicity.clusters.as.expected}. These simulations confirm both the hierarchical growth pattern and the prediction that the resulting accretion disk is magnetically-dominated and flux-frozen. More general models of ``supra-exponential'' or rapid seed growth \citep[see references above and e.g.][]{alexander:2014.rapid.growth,inayoshi:2016.hyper.eddington,inayoshi:2020.bh.seed.review,toyouchi:2021.super.eddington.imbh.growth.dusty.nuclear.disks.sims,su:2024.imbh.jet.selfreg.supereddington.accretion.challenging.idealized.flow,zana:2025.supereddington.accretion.protogalaxies.highz} share key features with this mode, although the scalings derived here differ in important details from those which assume Bondi-like or $\beta \gg 1$ accretion flows.

This ``leapfrog'' growth is inherently stochastic: it requires the BH to find itself embedded in a sufficiently dense cloud complex, which is rare but not implausible in the environments where the most massive star clusters and proto-bulges form. The duty cycle is low, but the growth rate during active episodes is so high that the long-term averaged growth can greatly exceed steady-state models. The end state of this phase --- a hyper-Eddington, low-mass BH ($M_{\rm BH} \sim 10^{5-6}\,M_{\odot}$) surrounded by a compact, magnetically-dominated accretion disk embedded in a dense stellar/gaseous envelope --- is remarkably similar to the properties inferred for the ``little red dots'' (LRDs) recently discovered by JWST at $z \gtrsim 4$ \citep[][and references therein]{matthee:2024.little.red.dots,greene:2024.lrd.unresolved,sneppen:2026.lrd.detailed.envelope.models.dynamical.masses.bhs.but.tiny.envelope.masses.extended.inflow.and.outflow.not.eqm,perezgonzalez:2026.lrd.many.flavors.with.different.spectra.variation.in.sfr.and.uv.and.ir,matthee:2026.lrd.spectra.strongly.shaped.column.densities.of.envelopes.obs,lin:2026.lrd.variability.confirmed.shows.expected.for.bh.with.envelope.confirms.agn.like,billand:2026.lrd.not.a.distinct.class.but.extreme.tail.of.continuous.distribution.of.galaxies.vshape.and.compact.correlate.continuously.classical.blr.and.dust.sufficient}. These objects show compact morphologies, broad emission lines implying BH masses of $\sim 10^{6}\,M_{\odot}$, and spectral energy distributions consistent with heavily reprocessed AGN accreting inside of ``envelopes'' of dense gas at highly super-critical ($\dot{m} \gg 1$) rates, in low-mass/faint host galaxies (references above). The ``leapfrog'' mode predicts precisely such objects as a natural outcome: the BH grows rapidly within the dense cloud complex that eventually becomes a compact proto-bulge, with the surrounding gas producing the observed ``envelope'' reprocessing and the magnetically-dominated disk powering the accretion (and therefore bolometric luminosity) while supporting the large covering factors (as predicted in \citealt{shi:2024.seed.to.smbh.case.study.subcluster.merging.pairing.fluxfrozen.disk}). 

\subsection{The Quasar Epoch: Massive BHs in Gas-Rich Galaxies}
\label{sec:cosmo:quasar}

For more massive BHs ($M_{\rm BH} \sim 10^{6-9}\,M_{\odot}$) in gas-rich galaxies at $z \sim 1-3$, the BH encounters clouds of comparable mass. This corresponds to the peak epoch of quasar activity \citep[e.g.][]{shen:bolometric.qlf.update,shen:2020.bolometric.qlf,hopkins:lifetimes.letter}, occurring in massive ($M_{\rm gal} \sim 10^{10-11}\,M_{\odot}$, $M_{\rm halo} \sim 10^{12} - 10^{13}\,M_{\odot}$), gas-rich galaxies with high nuclear gas surface densities ($\Sigma_{\rm cl} \sim 100-1000\,{\rm M_{\odot}\,pc^{-2}}$). Whether the compact, high-$f_{\rm gas}$ nuclear conditions are triggered by galaxy mergers, secular bar-driven inflows, or simply the higher cosmic gas accretion rates at early times \citep{hopkins:2006.qso.all,hickox:2018.agn.host.review,alexander:2025.bh.growth.agn.review}, the result is a population of GMC complexes with $M_{\rm cl} \sim 10^{7-9}\,M_{\odot}$ in the nuclear regions, placing these encounters in the comparable-mass (\S~\ref{sec:comparable.mass}) regime. These clouds are directly observed \citep[e.g.][]{newman:z2.clump.winds,swinbank:2012.population.disks.q1.highz.clumps,livermore:2015.highz.clump.galaxies.internal.structure,tacconi:2018.molecular.gas.depletion.masses.update,zhu:2026.jades.clumps}, but also physically inevitable in this regime. Because the Toomre mass of the ISM scales as $M_{0} \sim \psi\,\pi\,\Sigma_{0}\,H_{0}^{2} \propto f_{\rm gas}^{3}\,M_{\rm gal}$ (for a $Q \sim 1$ disk), the condition $M_{\rm BH} \sim M_{\rm cl} \sim M_{0}$ for the most vigorous, comparable-mass accretion mode is naturally satisfied in gas-rich galaxies where $f_{\rm gas} \sim 0.3-0.5$ and $M_{\rm gal} \sim 10^{10-11}\,M_{\odot}$ --- typical quasar hosts at $z \sim 2$ \citep[see discussion in][]{fukuda:nuclear.ring,thompson:rad.pressure,bournaud:2011.agn.fueling.by.clumps,forbes:2011.thick.disk.torque.evol}.

In the comparable-mass regime, individual cloud capture events produce vigorous accretion ($\dot{m} \sim 1-10$) lasting $\Delta t_{\rm event} \sim 0.1-10\,$Myr, with high duty cycles ($f_{\rm events}$ approaching or exceeding unity if $M_{\rm BH} \lesssim M_{0}$). The accretion is dominated by the most massive clouds, which also produce the most coherent magnetic fields and the thickest initial disks. This is consistent with the picture from the FORGE'd in FIRE simulations \citep{hopkins:superzoom.overview,hopkins:superzoom.disk,hopkins:superzoom.analytic,hopkins:multiphase.mag.dom.disks}, where cloud capture events from cosmological initial conditions produce magnetically-dominated, geometrically-thick accretion disks with $\beta \ll 1$ and $H/R \sim 0.1-1$.
This naturally connects the observed ``co-evolution'' of quasar accretion over cosmic time, ``downsizing'' as more-massive BHs tend to be in more gas-poor galaxies so begin to fall out of this regime and into the $M_{\rm BH} \gg M_{\rm cl}$ regime, clustering of and host galaxy properties of luminous quasars \citep{myers:clustering,shen:sdss.qso.clustering.update,shen:2012.qso.gal.crosscorr}, and the observed strong connection between AGN luminosity and host galaxy star formation rates \citep{hickox:2018.agn.host.review} and nuclear gas content \citep{medling:2021.agn.nuclear.gas,ho:2024.gas.content.agn}. The scaling of BH accretion rates with $\Sigma_{\rm cl}^{3/4}$ predicted here naturally explains why quasar activity peaks in the epoch when galaxy gas fractions and surface densities are highest.

\subsection{Low-Luminosity AGN in Massive Ellipticals}
\label{sec:cosmo:llagn}

For the most massive BHs ($M_{\rm BH} \sim 10^{8-10}\,M_{\odot}$) in quenched, gas-poor elliptical galaxies, the BH mass greatly exceeds typical cloud masses. Since $M_{0} \propto f_{\rm gas}^{3}\,M_{\rm gal}$, galaxies with $f_{\rm gas} \ll 0.1$ have $M_{0} \ll M_{\rm BH}$, placing the system firmly in the $M_{\rm BH} \gg M_{\rm cl}$ regime (\S~\ref{sec:highmassBH}). This corresponds naturally to the population of massive, bulge-dominated, quenched galaxies that dominate at $z \lesssim 1$ \citep{kormendy:2013.bh.review,mcconnell:2013.bhsigma}.
In this regime (studied in e.g.\ \citealt{cho:2023.multiscale.accretion.sims.bondi.inflow.model,cho:2024.multizone.grmhd.sims.bondi.flow.lowaccrate,cho:2025.bridging.scales.jet.host.galaxy.subgrid.model,su:2024.fire.jet.sim.using.acc.jet.prescriptions.from.cho.multiscale.experiments,su:2025.subgrid.model.suppressed.bondi.lowers.bh.growth.harder.to.rapidly.build.smbhs,porras.valverde:2025.suppressed.bondi.subgrid.prescription.in.sims.lowers.bh.growth.low.z.in.models,guo:2024.fluxfrozen.disks.lowmdot.ellipticals}), $\dot{m} \ll 1$, the field coherence scale $\sim R_{\rm cl} \ll r_{t}$, the depletion time $\Delta t_{d}$ can exceed $\Delta t_{\rm Parker}$, and the disk can thin sufficiently for the system to transition to a radiatively-inefficient accretion flow (RIAF/ADAF). Thus this naturally explains why the most massive BHs in quiescent galaxies are very rarely accreting at high Eddington ratios \citep[e.g.][]{ho:2009.llagn.review}. Indeed, in this regime, cosmological BH mass growth is believed to be increasingly dominated by BH-BH mergers following galaxy mergers, rather than by gas accretion \citep{shankar:2009.smbh.demographics.review}.

In these systems, Bondi-like accretion from a hot, virialized atmosphere may contribute, with the hot gas undergoing thermal instability to produce cold clumps \citep{gaspari:2013.chaotic.cold.accretion,gaspari:2017.cca.raining,li.bryan:2014.cooling.flow.cold.gas,tremblay:2016.m87.cold.gas,olivares:2019.cold.gas.clusters}, creating a hybrid of the ``cold cloud'' and ``hot Bondi'' accretion modes. Multi-scale simulations of this process \citep{cho:2023.multiscale.accretion.sims.bondi.inflow.model,cho:2024.multizone.grmhd.sims.bondi.flow.lowaccrate,guo:2024.fluxfrozen.disks.lowmdot.ellipticals,guo:2025.toroidal.disk,wang:2025.hypermagnetized.circumbinary.disk.flux.frozen.cavity.to.pc.scales} show precisely this behavior, transitioning from $\beta \gg 1$ Bondi flow to $\beta \ll 1$ cold disk and back to a quasi-spherical flow at small radii.

\section{Conclusions}
\label{sec:conclusions}

We have derived simple analytic scalings for the ``initial conditions'' of accretion disks formed by tidal capture and disruption of ISM clouds by black holes. The key conclusions are:

\begin{enumerate}[leftmargin=*]

\item{\textit{Cloud accretion can fuel BHs across all mass scales.} From stellar-mass seeds to the most massive SMBHs, the cloud accretion process operates in qualitatively different but well-defined regimes. For low-mass BHs ($M_{\rm BH} \ll M_{\rm cl}$), the dominant modes are turbulent Bondi-Hoyle capture (which is generally inefficient) and hierarchical ``leapfrog'' growth via sub-cloud encounters (which can be hyper-Eddington). For comparable-mass and high-mass BHs, individual cloud encounters produce vigorous accretion at or above the Eddington rate.}

\item{\textit{Clouds of mass $M_{\rm cl} \sim M_{\rm BH}$ dominate the accretion budget.} The time-averaged accretion rate, encounter rate, and duty cycle are all dominated by clouds of mass comparable to or modestly larger than the BH. This arises from the strong dependence of the capture radius and captured mass on $M_{\rm cl}/M_{\rm BH}$, combined with the cloud mass function.}

\item{\textit{The outer disk forms with $\beta \ll 1$, toroidally-magnetically dominated.} In all regimes where ISM clouds with $\beta_{\rm cl} \ll 1$ are accreted, the resulting disk at the circularization radius has $|B_{\phi}| \gtrsim |B_{z}|,\,|B_{R}|$ (the toroidal field is amplified by azimuthal stretching and vertical compression under flux-freezing), with $\beta_{d} \lesssim \beta_{\rm cl} \ll 1$. The turbulence is trans-\Alf{ic} and the thermal pressure is small. The disk is gravitationally stable with $Q \gg 1$.}

\item{\textit{The initial disk is geometrically thick, $H/R \sim 0.1-1$.} For the accretion events that matter (i.e.\ that produce significant accretion relative to the BH mass), the initial disk at the circularization radius is geometrically thick, with $H/R$ set by the \Alf\ speed rather than the thermal sound speed. This is always much thicker than a thermal or $Q \sim 1$ disk. For increasingly small clouds ($M_{\rm cl} \ll M_{\rm BH}$), $H/R$ decreases as $(M_{\rm BH}/M_{\rm cl})^{-1/6}$, but this simultaneously corresponds to $\dot{m} \ll 1$ and the transition to radiatively-inefficient accretion, which transitions back to quasi-spherical accretion for different physical reasons.}

\item{\textit{Accretion rate scalings.} For clouds with $M_{\rm cl} \lesssim M_{\rm BH}$, $\dot{M}_{d} \propto M_{\rm cl}^{3/4}$; for $M_{\rm cl} \gtrsim M_{\rm BH}$, $\dot{M}_{d} \propto M_{\rm cl}^{-1/4}$ (weakly declining). In all cases, $\dot{M}$ scales as $\Sigma_{\rm cl}^{3/4}$, resembling a Shu-like isothermal collapse scaling \citep{shu:isothermal.sphere.collapse} rather than Bondi-like or $\alpha$-disk scalings.}

\item{\textit{Magnetic coherence is maintained for $M_{\rm cl} \gtrsim M_{\rm BH}$.} When $M_{\rm cl} \gtrsim M_{\rm BH}$, the field coherence length is $\lambda_{\rm coh} \sim R_{\rm cl} \sim r_{t}$, ensuring that turbulent dissipation and Parker instability losses are slower than the advective/accretion timescale. For $M_{\rm cl} \ll M_{\rm BH}$, $\lambda_{\rm coh} \sim R_{\rm cl} \ll r_{t}$, and the field becomes more vulnerable.}

\end{enumerate}

\subsection{Comparison to Other Models}

These predictions differ qualitatively from several commonly-invoked models for AGN fueling: we summarize these here. 

\textit{Chaotic cold accretion} (CCA; \citealt{gaspari:2013.chaotic.cold.accretion,gaspari:2017.cca.raining}): Although both involve some similar concepts, the CCA model envisages accretion from many small, incoherent cold clumps produced by thermal instability in hot halos, with angular momentum cancellation from the isotropic distribution of clump orbits, and negligible magnetic fields (even in the cold gas) producing a thermal-pressure-dominated, razor-thin \citet{shakurasunyaev73}-like disk. In contrast, our calculations predict that accretion is dominated by a few events involving clouds of mass $\sim M_{\rm BH}$, each with \textit{coherent} angular momentum (sufficient to spin up the BH), producing magnetically-dominated disks rather than turbulently-cancelling flows and pure-hydrodynamic thin disks. The dimensional scalings for $\dot{M}$, $M_{\rm capt}$, and duty cycles differ qualitatively (and quantitatively by orders-of-magnitude) from CCA predictions, which are constructed for the $\beta \gg 1$, many-small-clump limit. It  remains possible that CCA could be applicable for the cooling/accretion from a very large-scale hot halo into, say a galaxy or galaxy nucleus, scales where the system is much larger than any cloud and fields are weaker, with a model like those here ``picking up'' and applying on small scales when it comes to actual gravitational capture by the BH.

\textit{Bondi accretion}: Standard Bondi or Bondi-Hoyle models assume $\beta \gg 1$, spherical symmetry, and steady-state flow. The cloud accretion scalings here instead resemble a series of hierarchical, one-to-one growth events where angular momentum plays a central role. The dimensional scalings differ: $\dot{M}_{\rm Bondi} \propto M_{\rm BH}^{2}\,\rho_{0}/v_{e}^{3}$, while the cloud accretion scalings go as $\dot{M} \propto v_{\rm cl}^{3}/G$, more akin to isothermal collapse rates at the captured-mass scale.

\textit{Effective-$\alpha$ disk models}: A time-integrated, orbit-averaged effective viscous transport model from large scales would predict $\langle \dot{M} \rangle \sim \Delta v^{3}/(G\,Q)$ for a $Q \sim 1$ galactic disk, and if modeled on classic $\alpha$-disk models effectively \textit{assumes} $\beta\gg 1$. This gives a rough match to the \textit{total} time-integrated accretion rate (since the most massive clouds have $v_{\rm cl} \sim \Delta v$), but fails to capture the stochastic, episodic nature of individual cloud encounters, the peak accretion rates per event, or the dependence on BH mass and cloud properties. More importantly it does not generalize to $\beta \ll 1$ unless one goes to a completely different self-similar model like that in \citet{hopkins:superzoom.analytic} built \textit{specifically} for the types of toroidally-magnetized disks discussed here. 

\subsection{Future Work}

The modeling in this paper is (intentionally) extremely simplistic. Our goal is by no means to be complete, nor to include all physics, nor to represent all modes of AGN accretion. There is much to do. High-resolution, multi-physics numerical simulations are required to validate and improve upon the scalings presented here. It is important to see, in such simulations, how the transition between the ``large'' and ``small'' cloud regimes occurs, and to better understand the ``leapfrog'' regime (which requires fully-modeling the interactions of BHs in non-linear, turbulent, clumpy media). We also make very simple assumptions for the population of clouds, which could be greatly improved. Probably most important, our goal here was to build up some simple intuition for the magnetic ``initial conditions'' of disks formed from cloud capture/disruption. Nothing in our modeling tells us how magnetic flux would be transported through the disk, and therefore whether a disk that \textit{forms} in the toroidally-magnetically-dominated state will \textit{remain} in that state (and whether the inner disk must have the same state as the outer disk). Similarly, the outer $\dot{M}_{d}$ we estimate, in addition to being crude, can (and almost certainly will) differ significantly from the actual accretion rate into the horizon or ISCO. Our work here is intended primarily for intuition-building and motivation for future, more detailed calculations.

\begin{acknowledgements}
We thank Haiyang Wang for a number of enlightening discussions during the development of this manuscript. Support for PFH was provided by a Simons Investigator Grant. Generative AI (Claude Opus 4.8) was used in the editing and revision of this paper text. 
\end{acknowledgements}

\bibliographystyle{mn2e}
\bibliography{ms_extracted}

\end{document}